\documentclass[acmtog]{acmart}

\setcitestyle{square}

\usepackage{units}

\usepackage{booktabs} 
\usepackage[flushleft]{threeparttable}
\usepackage[ruled,vlined]{algorithm2e}
\usepackage{wrapfig}

\usepackage{url}

\usepackage{gensymb}

\newcommand{\figref}[1]{Fig.\ \ref{#1}}
\newcommand{\tableref}[1]{Table \ref{#1}}
\newcommand{\secref}[1]{Sec.\ \ref{#1}}
\newcommand{\equref}[1]{Eq.\ \eqref{#1}}

\newcommand{\transposeSign}{{\!\top\!}}
\newcommand{\tr}{\transposeSign}

\usepackage{dsfont}
\newcommand{\R}{\mathds{R}}

\newcommand{\norm}[1]{\left\lVert #1 \right\rVert}
\newcommand{\normltwo}[1]{\norm{#1}_{2}^{2}}

\setcopyright{rightsretained}

\acmYear{2018}
\copyrightyear{2018}

\author{Oliver Glauser}
 \affiliation{%
   \institution{ETH Zurich}
 }
\author{Daniele Panozzo}
 \affiliation{%
   \institution{New York University}
 }
\author{Otmar Hilliges}
 \affiliation{%
   \institution{ETH Zurich}
 }
\author{Olga Sorkine-Hornung}
 \affiliation{%
   \institution{ETH Zurich}
 }

\begin{document}

\title{Deformation Capture via Soft and Stretchable Sensor Arrays}

\renewcommand{\shortauthors}{Glauser et al.}

\begin{abstract}
We propose a hardware and software pipeline to fabricate flexible wearable sensors and  use them to capture deformations without line of sight.
Our first contribution is a low-cost fabrication pipeline to embed multiple aligned conductive layers with complex geometries into silicone compounds.
Overlapping conductive areas from separate layers form local capacitors that measure dense area changes.
Contrary to existing fabrication methods, the proposed technique only requires hardware that is readily available in modern fablabs.
While area measurements alone are not enough to reconstruct the full 3D deformation of a surface, they become sufficient when paired with a data-driven prior.
A novel semi-automatic tracking algorithm, based on an elastic surface geometry deformation, allows to capture ground-truth data with an optical mocap system, even under heavy occlusions or partially unobservable markers.
The resulting dataset is used to train a regressor based on deep neural networks, directly mapping the area readings to global positions of surface vertices.
We demonstrate the flexibility and accuracy of the proposed hardware and software in a series of controlled experiments, and  design a prototype of wearable wrist, elbow and biceps sensors, which do not require line-of-sight and can be worn below regular clothing.

\end{abstract}

\begin{teaserfigure}
 \centering
 \includegraphics[width=1\textwidth]{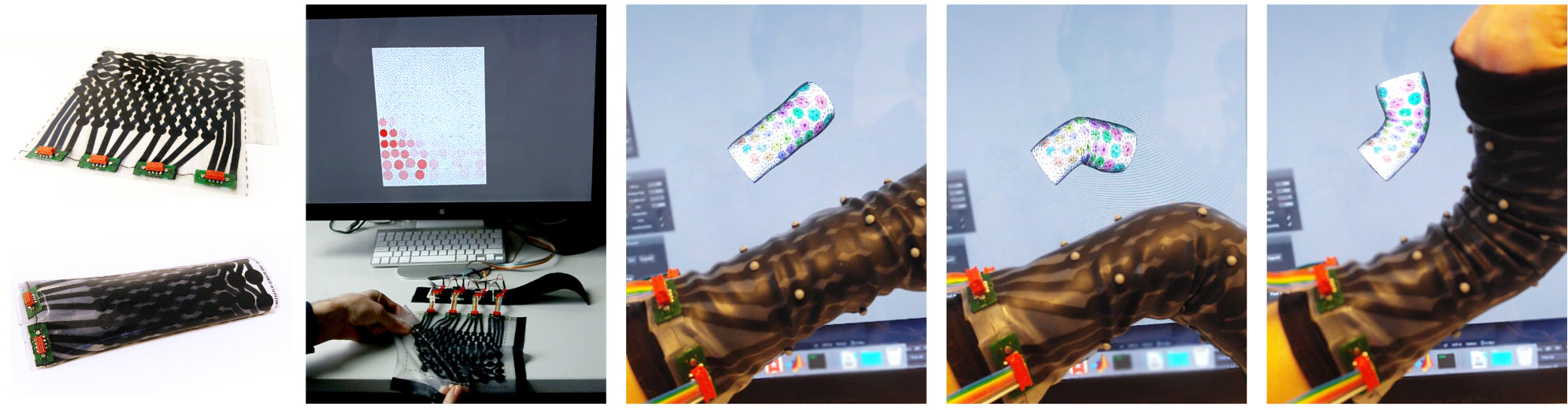}
 \caption{Left to right: We propose a method for the fabrication of soft and stretchable silicone based capacitive sensor arrays. The sensor provides dense stretch measurements that, together with a data-driven prior, allow for the capture of surface deformations in real-time and without the need for line-of-sight.}
 \label{fig:teaser}
\end{teaserfigure}

\maketitle

\section{Introduction}

Motion capture is an essential tool in many graphics applications, such as character animation for movies and games, sports, biomechanics, VR, and AR.
Most commonly, motion capture systems are camera based, either relying on body-worn markers or more recently markerless. Vision based approaches can be highly accurate and in the case of multiview or depth imaging, they can provide dense surface reconstructions. However, such systems rely on extensive infrastructure and are therefore mostly confined to lab and studio use.
Other sensing modalities, such as body-worn inertial and magnetic sensors, or resistive and capacitive distance sensors have been explored to provide more mobility, yet these are typically limited to capturing skeletal deformation only.

We introduce a new, practical and affordable approach to deformation sensing and motion capture. Our approach bridges the gap between vision-based and inertial approaches by providing accurate sensing of dense surface deformations while being wearable, and hence practical for scenarios in which stationary cameras are unsuited, for example to capture muscle bulging below clothing.

\paragraph{Capacitive sensor array}
We propose to leverage a capacitive sensor array, fabricated entirely from soft and stretchable silicone, that is capable of reconstructing its \emph{own} deformations.
The sensor array provides dense measurements of area change, which can be leveraged to reconstruct the underlying 3D surface deformation without requiring line-of-sight (see \figref{fig:no-line-of-sight-figure}). We furthermore contribute a data driven surface reconstruction technique, allowing for the capture of non-rigid deformations even in challenging conditions, such as under heavy occlusion, at night, outdoors, or for the acquisition of uncommon deformable objects.

Conductive polymers have been leveraged to fabricate resistive bend sensors \cite{Rendl12,Bacher16}, and are the basis of soft capacitive distance sensors, which are now readily available commercially \cite{StretchSense,ParkerHannifin}. Such stretchable capacitive sensors are enticing, since they are thin, durable, and may be embedded in clothing or directly worn on the body. However, so far fabrication has been involved and required specialized equipment, driving up cost. Moreover, such sensors have not been demonstrated to be accurate enough for motion capture and are typically limited to measurement of uniaxial deformation. Please note that capacitive sensing is often considered synonymous with touch sensing \cite{Grosse-Puppendahl:2017:FCG,Lee:1985:MTD:317456.317461,Rekimoto:2002:SIF}, in which capacitive coupling effects are leveraged to detect finger contact with a static sensor. In this paper, however, the term is used in a different sense, referring to the fact that capacitance changes when an electrode undergoes deformations.

\paragraph{Custom fabrication method}
We introduce a fabrication method for soft and stretchable capacitive deformation sensors, consisting of multiple bonded layers of conductive and non-conductive silicone. Crucially, the method only requires casting silicone and etching conductive traces by a standard laser cutter, and can thus be performed using hardware commonly available in a modern fabrication lab. The precision and accuracy of our sensors is comparable to commercial solutions, and the involved material costs are low.
Our approach supports embedding many sensor cells of custom shape in a single thin film. Each cell measures changes of its own area, caused by deformation of the surface it is attached to. The resulting sensor array can be read out at interactive rates.

\paragraph{Geometric prior}
While providing a rich signal, the area measurements alone are not sufficient to uniquely reconstruct the full 3D sensor shape due to isotropy and lack of direct bending measurement. They are however sufficient when paired with an appropriate geometric prior, if expected deformations involve some amount of non-area preserving stretch. 

In addition to the hardware, we propose an effective pipeline to acquire the deformation of the sensor worn by a user, for example wrapped around the wrist or an elbow. We propose a data driven technique based on a neural network regressor to reconstruct the sensor geometry from area measurements. At runtime, the regressor estimates the location of a sparse set of vertices, and the dense deformed surface is computed by a nonlinear elastic deformation method, obtaining a high-resolution reconstruction in real-time (see \figref{fig:teaser}).

To acquire the necessary training data, we overcome an additional challenge: optical tracking systems struggle with the heavy occlusions and large deformations typical for natural motions of wrists, elbows and other multi-axial joints. Furthermore, when capturing other non-rigidly deforming objects, skeletal priors cannot be leveraged to recover missing markers. We thus introduce a semiautomatic ground truth acquisition technique, enabling capture of the necessary training data in minutes and reducing tedious manual cleanup to a minimum. The approach leverages an elastic simulation of the sensor to disambiguate the marker tracks, deal with unlabeled markers and correctly attribute marker positions to the digital mesh model of the sensor.

\paragraph{Evaluation}
We demonstrate our sensors in action by acquiring dense deformations of a wrist and lower part of the hand (see \figref{fig:teaser}), an elbow, an inflating balloon, and muscle bulging. We also capture deformations of flat sensors, both in and out of plane, which shows the precision and localization properties of our capacitive sensor arrays. Finally, we evaluate the prediction accuracy of the learning based prior quantitatively.


\section{Related Work}
\label{sec:related}

Our work relates to several areas of the literature ranging from digital fabrication to motion capture and self-sensing input devices. We briefly review the most important work in these areas.

\subsubsection*{Camera-based motion capture.}
The acquisition of articulated human motion using cameras is widely used in graphics and other application domains. Commercial solutions require wearing marker suits or gloves and depend on multiple calibrated cameras mounted in the environment. To overcome these constraints, research has proposed marker-less approaches using multiple cameras (cf.\ \cite{moeslund2006survey}); sometimes these rely on offline \cite{bregler1998tracking,ballan2012motion,starck2003model} and more recently online processing \cite{rhodin2015versatile,deAguiar08,stoll2011fast,elhayek2017marconi}, but always require fixed camera installations.
Neumann et al al. \shortcite{Neumann2013} capture muscle deformations of a human shoulder and arm with a multi-camera system and derive a data-driven statistical model. 

\begin{figure}
	\centering
	\includegraphics[width=1.0\linewidth]{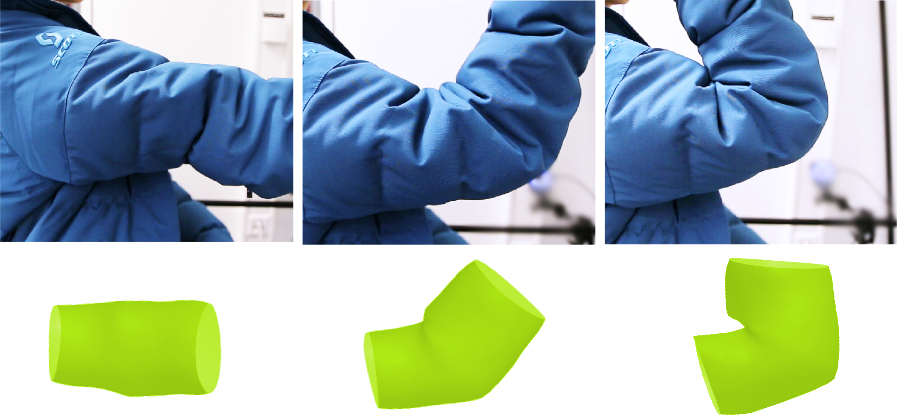}
	\caption{ An elbow ``hidden'' below by a jacket. Top: Video frames for comparison. Bottom: With our approach the dense surface deformation is estimated without requiring line of sight.}
	\label{fig:no-line-of-sight-figure}
\end{figure}
Recent pose estimation methods exploit deep convolutional networks for body-part detection in single, fully unconstrained images~\cite{chen2014articulated,newell2016stacked,tompson2014joint,toshev2014deeppose,wei2016cpm}. However, these methods only capture 2D skeletal information. Predicting 3D poses directly from 2D RGB images has been demonstrated using offline  methods \cite{bogo2016keep,tekin2016fusing,zhou2016sparseness} and in online settings \cite{VNect_SIGGRAPH2017}. Monocular \emph{depth} cameras provide additional information and have been shown to aid robust skeletal tracking \cite{Ganapathi2012real,ma2014real,taylor2012vitruvian,shotton2013real,Taylor:2016:EPI} and enable dense surface reconstruction even under deformation \cite{zollhofer2014real,newcombe2015dynamicfusion,Dou:2016:FRP}. Multiple, specialized structured light scanners can be used to capture high-fidelity dense surface reconstructions of humans \cite{Pons-Moll15}.

All vision-based approaches struggle with visual clutter, (self-)occlusions and difficult lighting conditions, such as bright sunshine in the case of depth cameras, high contrast or lack of illumination in the case of color cameras. Furthermore, all camera based systems require line-of-sight and often precise calibration, and are therefore not well suited in many scenarios, such as outdoors. Our sensor is a first step in removing these limitations, allowing mobile and self-contained sensing, without line of sight.

\subsubsection*{Self-sensing input devices.}
An important feature of our method is the capability of measuring the sensor's own deformation without requiring any external cameras. Such \emph{self-sensing} input devices, usually not designed for motion capture, have  been first demonstrated in the Gummi system \cite{Schwesig2004}, which simulated a handheld, flexible display via two resistive pressure sensors. Other early work used the ShapeTape sensor \cite{Danisch1999} for input into a 3D modeling application~\cite{Balakrishnan1999}. 
Metallic strain gauges embedded into flexible 3D printed 1D strips measure the bending and flexing of custom input devices \cite{Chien15}. Rendl et al.~\shortcite{Rendl14} use eight transparent printed electrodes on a transparent and flexible 2D display overlay to reconstruct $2.5$D bending and flexing of the sheet in real time, but do not allow for stretch. \cite{Bacher16} propose an optimization based algorithm to design self-sensing input devices by embedding piezo-resistive polymer traces into flexible 3D printed objects. \cite{Sarware17} use polyacrylamide electrodes embedded in silicone to produce a flexible, transparent 4$\times$4 sensing grid, and \cite{Xu16} propose a PDMS based capacitive array; both are limited to detecting touch gestures. Hall effect sensors embedded into hot-pluggable and modular joints can measure joint angles of tangible input devices used for character animation \cite{Jacobson:TMIDCA:2014,Glauser:2016:ATMID}. While demonstrating the rich interactive possibilities afforded by flexible input devices, none of the above approaches are directly suitable for the acquisition of dense non-rigid surface deformation.

\subsubsection*{Inertial measurement units (IMUs).}
Attaching sensors directly onto the body overcomes the need for line-of-sight and enables use without infrastructure. IMUs are the most prominent type of sensors used for pose estimation. Commercial systems rely on 17 or more IMUs, which fully constrain the pose space, to attain accurate \emph{skeletal} reconstructions via inverse kinematics \cite{roetenberg2007moven}. Good performance can be achieved with fewer sensors by exploiting data-driven methods \cite{tautges2011motion,liu2011realtime,schwarz2009discriminative} or taking temporal consistency into account, albeit at high computational cost and therefore requiring offline processing \cite{vonMarcard17}. While IMUs provide mobility and accuracy, they cannot sense dense surface deformations.

\subsubsection*{Strain gauges, stretch and bend sensors.}
Strain sensors fabricated from stretchable silicone and attached directly to the skin have been proposed to measure rotation angles of individual joints~\cite{Lee16}. Shyr et al.~\shortcite{Shyr14} propose a textile strain sensor, made from elastic conductive yarn, to acquire bending angles of elbow and knee movements. \cite{Mattmann08} and \cite{Lorussi04} use strain gauges embedded into garments to classify discrete body postures. \cite{Scilingo03} propose polymerized fabric strain sensors and demonstrate use of the sensor in a data glove. Specifically designed for the capture of wrist motion, \cite{Huang17} use five dielectric elastomer sensors and achieve an accuracy of 5\degree\ for all motion components, highlighting the difficulty of reconstructing joint orientation of complex, multi-axial joints such as the wrist, shoulder or ankle.
Bending information can be used to recover articulated skeletal motion, and resistive bend sensors are typically used in VR data gloves. However, these suffer from hysteresis \cite{Bacher16}; imprecise placement and sensor slippage can impact accuracy \cite{kessler1995evaluation}. A soft bend sensor that is insensitive to stretching and mountable directly on the user's skin is proposed in \cite{Shen16}, increasing angular accuracy, but it is inherently limited to measuring uni-axial bending.

We propose a wearable, soft and stretchable silicone-based capacitive sensor design, focused on measuring dense area changes, which allows us, in combination with a data-driven reconstruction technique, to accurately capture dense, articulated and non-rigid deformations.

\subsubsection*{Fabrication.} Producing capacitive elastomer stretching sensors is challenging, and the mechanical, electrical and thermal properties all depend on the type of material used and the pattern of conductive traces or electrodes. Another challenge is that the silicone is hydrophobic, hence the adhesion of non-silicones is extremely difficult. For an extensive review of various ways to manufacture conductive layers for such sensors or actuators, we refer to \cite{Rosset13}. Composites of carbon black (conductive powder) and silicone are widely used, see e.g.\ \cite{Araromi15, Rosset16, Huang17, OBrien14}. A large range of fabrication methods for manufacturing conductive trace patterns have been proposed. Most methods rely on the potentially costly fabrication of intermediate tools like screen printing masks \cite{Jeong16, Wessely16}, molds \cite{Huang17, Sarware17} or stencils \cite{Rosset16}. To circumvent the adhesion issue, specialized plasma chambers are often required to selectively pre-treat the base layer \cite{Jin17}. An alternative procedure, introduced by \cite{Lu14}, involves patterning conductive PDMS sheets, manually removing excess parts with tweezers, sealing the resulting circuit with PDMS and bonding multiple such circuit layers to form capacitive touch sensors (as demonstrated by \cite{Weigel15}). Similar to \cite{Araromi15}, our  process leverages a standard laser cutter to etch away the negative sensor pattern, opening up the possibility to digitally design electrode patterns and produce them with low error tolerance. 
	However, in contrast to prior work, our fabrication method does not require a plasma chamber or manual alignment and gluing of the different layers. Hence it allows for the production of larger sensors with a high alignment quality (see \figref{fig:alignment}). To the best of our knowledge, we are the first to propose a fabrication method that requires almost no specialized hardware and enables creating large high-resolution multi-layer sensor \emph{arrays}.

\subsubsection*{Capacitive (touch) sensing}
Ever since the introduction of the Theremin \cite{glinsky2000theremin}, an experimental musical instrument, researchers have explored the use of capacitive sensing in the context of HCI. Most notably, capacitive coupling effects are the basis of early \cite{beck1973two,Lee:1985:MTD:317456.317461} and virtually all modern touchscreen devices \cite{Rekimoto:2002:SIF}. Capacitive coupling effects exist naturally between many objects (including humans) and their surroundings, and by measuring the changes in relative values it is possible to recover relative position, proximity and other properties. The seminal works by Smith \shortcite{smith1995toward} and Zimmermann et al.~\shortcite{Zimmerman:1995:AEF} introduced and categorized the various electric field sensing aspects to the interaction research community and demonstrated applications that went well beyond binary touch detection. Since then capacitive coupling effects have been used to sense touch, detect and discriminate user grip and grasp, detect and track objects on interactive surfaces, track 3D positions and proximity and coarsely classify 3D poses and gestures. We refer to the survey by Grosse-Puppendahl et al.~\shortcite{Grosse-Puppendahl:2017:FCG} for an exhaustive treatment. Notably, flexible and bendable sensors \cite{Gotsch:2016:HFH:2851581.2890258,Han:2014:TDE:2642918.2647381,Poupyrev:2016:PJI:2858036.2858176} and those directly worn on the user's skin \cite{Weigel15,kao2016duoskin,Nittala2018} have been proposed. However, virtually all of the above work measures one or a combination of different capacitive coupling effects, that is, the change in capacitance due to a conductive object (such as a finger) approaching an electrode. Our work is fundamentally different in that we do not sense capacitive coupling effects but instead measure changes in the electrodes' properties themselves: under deformation,  the area of the electrode's plates changes, which in turn changes the capacitance of the plate and hence the charge time of the capacitor. We show how this effect can be leveraged to recover, using appropriate geometric priors, detailed 3D surface deformations, albeit at the cost of requiring a custom read-out scheme.

\section{Overview}
We present a stretchable silicone elastomer based sensor and its corresponding fabrication procedure. The sensor senses its \emph{own deformation} and estimates the local surface area changes during deformation when wrapped around an object or a body part of interest (e.g., a wrist). The sensor array is fabricated layer onto layer entirely from 2-component silicone elastomer with conductive elements made from the same silicone but mixed with carbon black particles. The conductive layers can be designed to contain custom electrode patterns via etching with a standard laser cutter. This approach avoids the production of masks or molds and makes interlayer alignment very straightforward and precise. 

As a further contribution we introduce a silicone-based capacitive area sensor \emph{array}, whereas prior work only demonstrated individual stretch sensing elements, and arrays only to detect dense touch or pressure (e.g.,  \cite{Lipomi2011,Sarware17,Nittala2018, Engel2006,Wong2012,Wissman2013,Block2013,Woo2014}).

Our key insight is that such arrays could also be used to attain dense localized area changes, given an appropriate read-out scheme. Our arrays are made by placing electrode strips in two conductive layers, separated by a dielectric, together forming a non-uniform grid of capacitors.
Furthermore, we propose a scanning based read-out scheme that does not require individually connected capacitors, which would require a large number of layers or a large portion of the sensor area dedicated to connection leads. Instead, we propose a time-multiplexing procedure to indirectly read out capacitance values, which allows for a drastically simplified routing of electric connections.  By integrating all the capacitance readings, we can acquire area changes with a sufficient granularity and accuracy to reconstruct the geometry of an object, given suitable geometric priors. These dense area measurements are therefore combined with a deep learning based regressor to attain 3D position estimates of key points on the surface and an elastic deformation optimization to obtain dense deformation reconstructions.

In the following sections we provide a brief primer on capacitive sensing (\secref{sec:principles}), detail our sensor design (\secref{sec:array}) and detail the fabrication (\secref{sec:fabrication}). We then complete our method by introducing our data capture and cleanup, learning and surface reconstruction approaches (\secref{sec:deformation}).

\section{Sensor Design}
\label{sec:sensordesign}

\subsection{Preliminaries}
\label{sec:principles}
The capacitance $C$ (in Farads) of a plate capacitor is given by
\begin{equation}
C = \epsilon_r \epsilon_0 \frac{A}{d} = \epsilon_r \epsilon_0 \frac{lw}{d},
\label{eqn:classicCap}
\end{equation}
where $A$ is the area of overlap of the two electrodes (in square meters), $\epsilon_r$ is the
dielectric constant, $\epsilon_0 $ is the electric constant and $d$ is the separation between the plates (in meters). Assuming a rectangular plate capacitor, $l$ is its length and $w$ the width.
\setlength{\intextsep}{5pt}%
\setlength{\columnsep}{5pt}%
\begin{wrapfigure}{r}{0.45\linewidth} 
    \includegraphics[width=\linewidth]{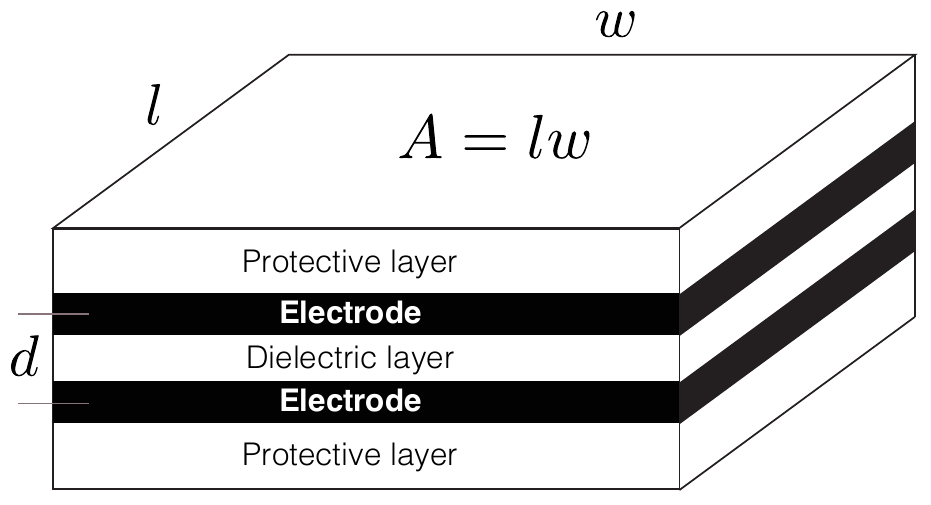}
\end{wrapfigure}
While originally derived for static plate capacitors, this relationship also holds for capacitors made from silicone elastomers \cite{Atalay17,Huang17,OBrien14}. To minimize capacitive coupling effects with other objects, capacitors are typically shielded via insulating layers
(see inset). Using \equref{eqn:classicCap}, and assuming the same Poisson ratio of width and thickness of the sensor ($d/d^0=w/w^0$), a linear relationship between the ratio of the stretched capacitor's length $l$ to the rest pose length $l^0$, and the ratio of the capacitance of the stretched capacitor $C$ to the rest pose capacitance $C^0$ can be established:
\begin{equation}
\frac{C}{C^0} = \frac{\epsilon_r \epsilon_0 \frac{lw}{d}}{\epsilon_r \epsilon_0 \frac{l^0 w^0}{d^0}} = \frac{l}{l^0} \frac{w}{w^0} \frac{d^0}{d} = \frac{l}{l^0}.
\label{eqn:classicDist}
\end{equation}
Prior work applies this principle to the design of capacitive, uni-axial stretch sensors \cite{Atalay17} by continuously measuring a capacitance, which is then transformed to length measurement using  \equref{eqn:classicDist}. Note that here, an assumption is made that stretch only happens along $l$, which typically requires fabricating isolated, individual capacitors (\figref{fig:concept}a). Our aim is to create a dense array of sensing elements, for which stretch may occur in multiple directions and hence each sensing element captures changes in area.

\subsubsection*{Area changes.} Starting from  \equref{eqn:classicCap}, and assuming volume conservation ($V = V^0 \ \Leftrightarrow \ A \,d = A^0 d^0 \ \Leftrightarrow \ d^0/d = A/A^0$) and constant stretch throughout the entire sensor cell, the ratio of capacitance before and after deformation can be expressed as
\begin{equation}
\frac{C}{C^0} = \frac{\epsilon_r \epsilon_0 \frac{A}{d}}{\epsilon_r \epsilon_0 \frac{A^0}{d^0}} = \frac{A}{A^0} \frac{d^0}{d} = \left(\frac{A}{A^0}\right)^2.
\label{eqn:cap2area}
\end{equation}
Thus, if we know the current capacitance $C$ of a sensor cell and have recorded its rest pose area $A^0$ and capacitance $C^0$, we can compute the change in area between the rest state and the current configuration as
\begin{equation}
\frac{A}{A^0} = \sqrt{\frac{C}{C^0}}.
\label{eqn:area_to_capacitance}
\end{equation}

\paragraph{Touch vs. pressure vs. stretch.} We note that there are fundamental differences between capacitive sensing of touch, pressure, and stretch. The majority of the HCI literature on capacitive sensing measure capacitive coupling effects (e.g., changes in capacitance due to an approaching finger). Applied pressure can be measured capacitively since the thickness $d$ is reduced, which leads to a higher capacitance $C$ (see \equref{eqn:classicCap}). Finally, in our work, both the overlap area $A$ and the thickness $d$ change due to the deformation of the sensor, requiring a custom read-out scheme (cf.\  \figref{fig:ghosting}). We now explain how a naive implementation, designed for touch or pressure sensing, must be modified in order to capacitively sense deformation.

\subsection{Sensor layout}
\label{sec:array}
\begin{figure}
	\centering
	\includegraphics[width=1\linewidth]{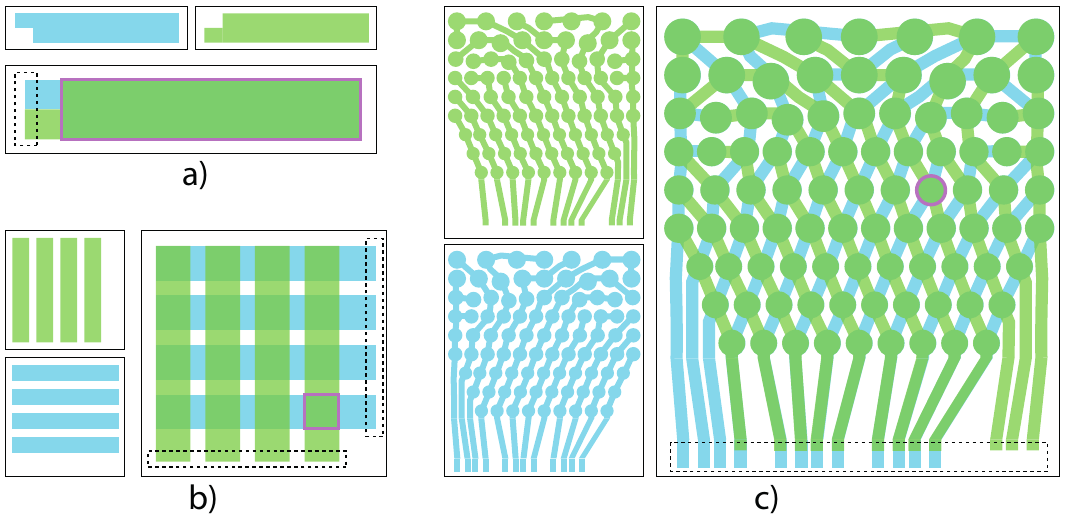}
	\caption{Various electrode strip patterns, with the bottom layer in blue and the top layer in green. When overlaid, the overlapping regions form \emph{sensor cells}; we highlight one cell in each example in pink. The dashed lines outline the places where the read-out circuit is connected. Example (a) is a classic elastomer strain sensor with 2 leads and 1 sensor cell; (b) is our array concept with 8 leads and 16 sensor cells; (c) depicts our actual prototype sensor, a warped grid that brings all connection leads to the bottom side of the sensor, with 24 leads and 92 sensor cells.}
	\label{fig:concept}
\end{figure}

\begin{figure}
	\centering
	\includegraphics[width=1\linewidth]{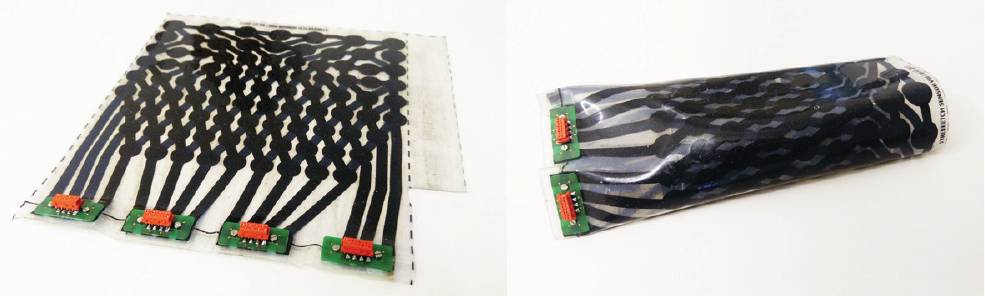}
	\caption{Left: Our prototype sensor with connector boards. Both conductive layers contain 12 electrode strips each, and the overlaps amount to 92 sensor cells. Right: Using silicone glue, the topology of a flat sensor can be changed to form e.g.\ a cylinder. See \figref{fig:bicepslayoutphoto} for our second and larger fabricated sensor.}
	\label{fig:sensor-flat-cylinder}
\end{figure}

Dense surface deformation capture requires a sensor that can measure local changes in the surface geometry with high density. This need has to be balanced with the complexity of the electrical design, so that the fabrication remains feasible.
Our proposed concept of the sensor array (\figref{fig:concept}b), which we call simply \emph{sensor} from now on, strikes this balance with its two-electrode-layers design. 
The sensor is made of two conductive layers with $n$ and $k$ independent electrode strips on each layer, respectively. We call the individual electrodes \emph{strips}, but they may have any shape.
Overlapping sections of two electrode strips from separate layers form a local capacitor, which we call a \emph{sensor cell} $S$.
We lay out the strips in a non-uniform grid arrangement, as shown in \figref{fig:concept}c. Each pair of strips from top and bottom layers crosses at most once, amounting to $s$ sensor cells ($s \leq k  n$). This design allows routing all strips to the same side of the sensor, where the silicone-based traces are connected to a PCB for the measurement of capacitances (\figref{fig:sensor-flat-cylinder}). However, since sensor cells are daisy-chained, we cannot directly read each one independently. We now derive a read-out scheme that provides the desired localized area measurements.

\begin{figure}
	\centering
	\includegraphics[width=1\linewidth]{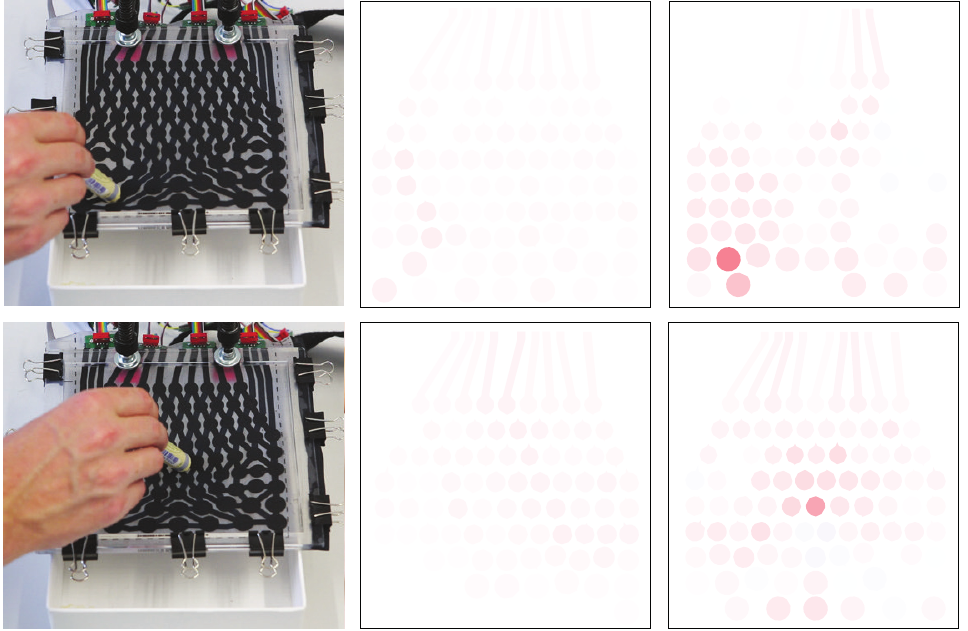}
	\caption{A naive scanning scheme (mutual-capacitance approach, using charging time to measure capacitance) results in underestimation of the magnitude of stretch, leads to not well-localized measurements, and even gives incorrect readings. Left: Sensor is deformed by poking with a pen. Middle: Change of magnitude per sensor cell, measured by the naive scanning scheme. Right: Change of magnitude per sensor cell, measured by our proposed scheme (see the respective video clip in supplemental material).}
	\label{fig:ghosting}
\end{figure}

\begin{figure*}[t]
	\centering
	\includegraphics[width=1\linewidth]{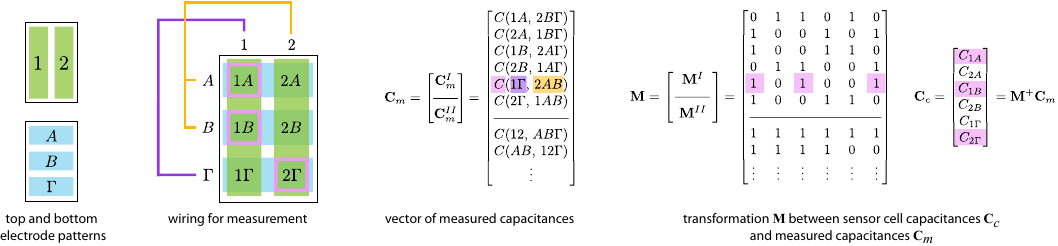}
	\caption{Measuring capacitance of sensor cells via selective combinations of strips. The measured combination in this example is comprised of strips $1$ and $\Gamma$ as the source electrode, and strips $2$, $A$ and $B$ as the ground electrode. The resulting overlaps are highlighted in pink. The measurement contributes the equation $C(1\Gamma,2AB) = C_{1A} + C_{1B} + C_{2\Gamma}$ to the linear system that recovers the individual sensor cell capacitances. }
	\label{fig:example3x3}
\end{figure*}

\paragraph{Sensor read-out.}\label{sec:readout}
As mentioned, our sensor is designed to consist of only two capacitive layers, which renders individual addressing of capacitors difficult without sacrificing sensor surface for complex routing of electrical traces. We experimentally verified that simple scanning schemes common in mutual capacitive touchscreens cannot be applied in the case of geometrically deforming and overlapping capacitor plates and traces, see \figref{fig:ghosting}.

We propose a time-multiplexing scheme, in which a voltage is applied to a subset of strips from both layers in turn, and the remaining strips are connected and serve as the second plate of the local capacitor.
A simple example of a sensor composed of a 3$\times$2 grid of electrode strips, with a total of $s = kn = 6$ sensor cells, is shown in \figref{fig:example3x3}.
For each such measurement, the cells where the combined electrode strips overlap are measured in parallel.
The capacitances of these cells add up, leading to a linear relationship between the individual sensor cell capacitances and the measured, combined capacitance. This can be expressed in matrix form:
$$ \mathbf{M} \, \mathbf{C}_c  = \mathbf{C}_m\,.$$
Here, $\mathbf{M}$ is an $s \times s$ binary matrix with rows encoding different measurement combinations, so that $\mathbf{M}$  transforms the vector of sensor cell capacitances $\mathbf{C}_c$ into the measured capacitances $\mathbf{C}_m$.
Using our example in \figref{fig:example3x3} to illustrate the composition of this linear system of equations, the vector $\mathbf{C}_c$ is
\begin{equation}
\mathbf{C}_c = \left[
		C_{1A},\
		C_{2A},\
        C_{1B},\
        C_{2B},\
        C_{1\Gamma},\
        C_{2\Gamma}\
       \right]^\tr,
\end{equation}
where $C_{1A}$ denotes the  sought localized capacitance of sensor cell $1A$, and so on.
Each row of $\mathbf{M}$ corresponds to a measurement, where the row elements corresponding to jointly read sensor cells are set to 1 and the remaining elements to 0.
In our example (\figref{fig:example3x3}), the highlighted row of $\mathbf{M}$ corresponds to a measurement where electrodes $1$ and $\Gamma$ are connected to serve as the source electrode, and $2, A, B$ as the ground electrode. This leads to cells $1A, 1B \text{ and } 2\Gamma$ to form parallel capacitors, and the read-out values are summed.

To reconstruct $\mathbf{C}_c$ from measurements $\mathbf{C}_m$, the matrix $\mathbf{M}$ needs to be invertible, which is the case if it has $s$ linearly independent rows. The matrix $\mathbf{M}^I$ is formed by iteratively connecting one strip from the top and bottom layer as source electrode, with all remaining strips connected as the ground electrode, resulting in the required $s$ linearly independent rows.
We experimentally found that taking additional measurements with all remaining combinations of strips, collected in matrix $\mathbf{M}^{II}$, and solving the resulting over-constrained linear system in the least square sense leads to extra robustness:

\begin{equation}
\mathbf{C}_c = \mathbf{M}^+ \mathbf{C}_m.
\label{eqn:sensor2circle}
\end{equation}
Here,
\begin{equation}
\mathbf{M} =  \begin{bmatrix}
		\mathbf{M}^I\\
        \textrm{\textemdash\textemdash} \\
        \mathbf{M}^{II}\\
		\end{bmatrix}, \ \
\mathbf{C}_m = \begin{bmatrix}
		\mathbf{C}_m^I\\
        \textrm{\textemdash\textemdash}\\
        \mathbf{C}_m^{II}\\
		\end{bmatrix}
,
\end{equation}
where $\mathbf{C}_m^{I}, \mathbf{C}_m^{II}$ represent the capacitance readings of the mandatory part $\mathbf{M}^I$ and the additional measurements $\mathbf{M}^{II}$, respectively.

\paragraph{Non-uniform stretch.} Since our sensor cells have non negligible size (\figref{fig:sensor-flat-cylinder}), the uniform stretch assumption may not hold in practice. We therefore model a sensor cell $S_j$ more accurately by splitting it into several elements (triangles) $e_i \in S_j$, each with an individual (uniform) area stretch. Applying \equref{eqn:cap2area} to each element, the capacitance $C_j$ of the sensor cell becomes
\begin{equation}
\frac{C_j}{C^0_j} = \frac{1}{C^0_j}\sum_{e_i \in S_j} \left(C^0_j \frac{{A_i^0}}{{A_j^0}}\right) \left(\frac{A_i}{A_i^0}\right)^2 = \frac{1}{A^0_j}\sum_{e_i \in S_j} \frac{{A_i}^2}{{A_i^0}},
\label{eqn:face2cap}
\end{equation}
where $C_i^0 = C^0_j {A_i^0}/{A_j^0}$ is the rest pose capacitance of element $e_i$. This holds because in rest state, the thickness $d$ is constant, and hence the rest state capacitance is proportional to the area $A_i^0$.

\subsection{Fabrication}
\label{sec:fabrication}

\begin{figure*}[t]
	\centering
	\includegraphics[width=\textwidth]{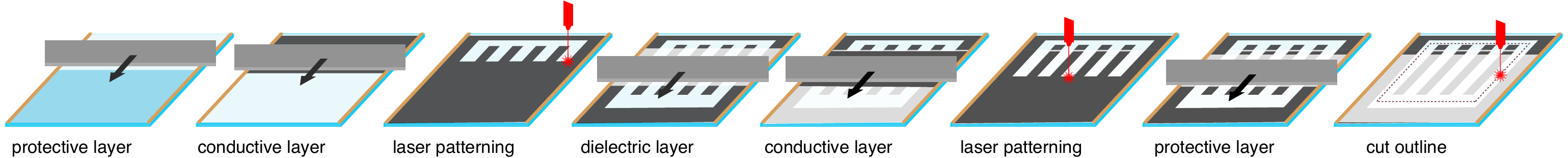}
	\caption{The proposed fabrication pipeline consists of eight main steps. From left to right: Casting a protective layer; casting a conductive layer; etching the negative electrode strip pattern with a laser cutter; dielectric layer; conductive layer; etching again; protective layer; cutting the desired outline.}
	\label{fig:fabpipeline}
\end{figure*}

We propose a fabrication pipeline, illustrated in \figref{fig:fabpipeline}, for silicone-based sensors with arbitrarily shaped electrodes. 

\subsubsection*{Structure.}
The sensor consists of two conductive layers with a dielectric layer between them, and it is encased by shielding layers (see inset on the previous page). During fabrication the sensor rests on a flat glass plate to which the silicone elastomer sticks well but the final sensor can be easily detached. We provide the  description of the chemical composition of the silicone mixtures in Appendix~\ref{sec:mixtures}. The layers are cast one by one by spreading the silicone using a blade; the correct thickness is ensured by Kapton tape (\unit[65]{$\mu$m} thickness) at the borders of the glass plate.
After the casting of each layer the sensor is cured for 20 minutes in an oven at $100\,^{\circ}\mathrm{C}$.

The second, conductive layer (silicone mixed with carbon black) is directly cast onto the shielding layer, and after curing, the desired pattern is etched with a laser cutter. The etching is done with a 100 Watt Trotec Speedy 360 laser cutter. Two rounds of etching are carried out with the following settings: 20 Power, 60 Speed and 500 Pulses/inch. This vaporizes the carbon black to create non-conductive areas between traces, while the underlying silicone-only layer stays intact. The resulting dust can be carefully removed with isopropyl alcohol without damaging the electrodes. The sensor is completed by adding another dielectric, the second capacitive layer (which is also etched and cleaned) and finally another shielding layer. The overall process takes around 3.5 hours (\unit[1]{h} for mixing and casting, \unit[1.5]{h} for curing and \unit[1]{h} for laser etching) for producing a sensor of 200$\times$\unit[200]{mm}.

\begin{figure}
	\centering
	\includegraphics[width=1\linewidth]{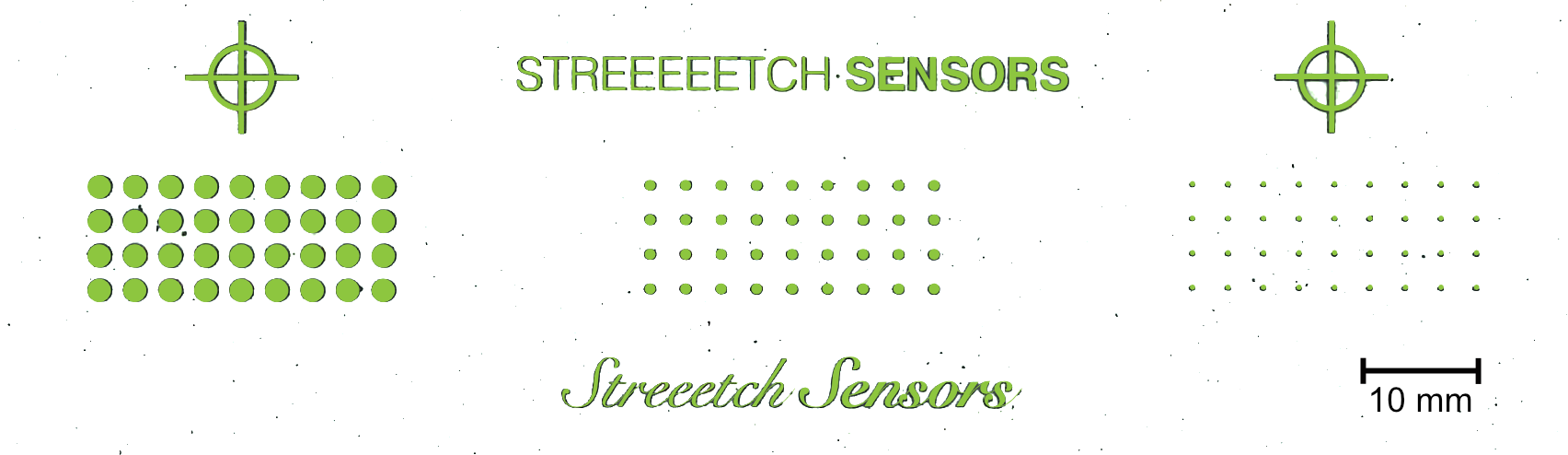}
	\caption{To demonstrate the alignment quality of our fabrication method, we produced a test pattern with two identical conductive (black) layers. The fabricated pattern was scanned with a flatbed scanner. The scan is overlaid with the digital design (green). Wherever the alignment is perfect, only the green layer is visible.}
	\label{fig:alignment}
\end{figure}

In previous works \cite{Lu14,Araromi15}, the alignment of the different layers of a multilayer sensor had to be done manually. Aligning the layers with high accuracy and without wrinkles can prove a difficult task, especially for larger sensors like ours. With our approach, a high alignment quality is achieved by design, since we directly cast layers onto one another (see the accompanying video from 01:05) and place the base glass plate in the laser cutter aligned with physical stoppers before etching. \figref{fig:alignment} shows an alignment experiment. 

The thickness of the final sensor is about \unit[500]{\micro m}, the conductive
\setlength{\intextsep}{3pt}%
\setlength{\columnsep}{8pt}%
\begin{wrapfigure}{r}{0.18\linewidth} 
    \includegraphics[width=\linewidth]{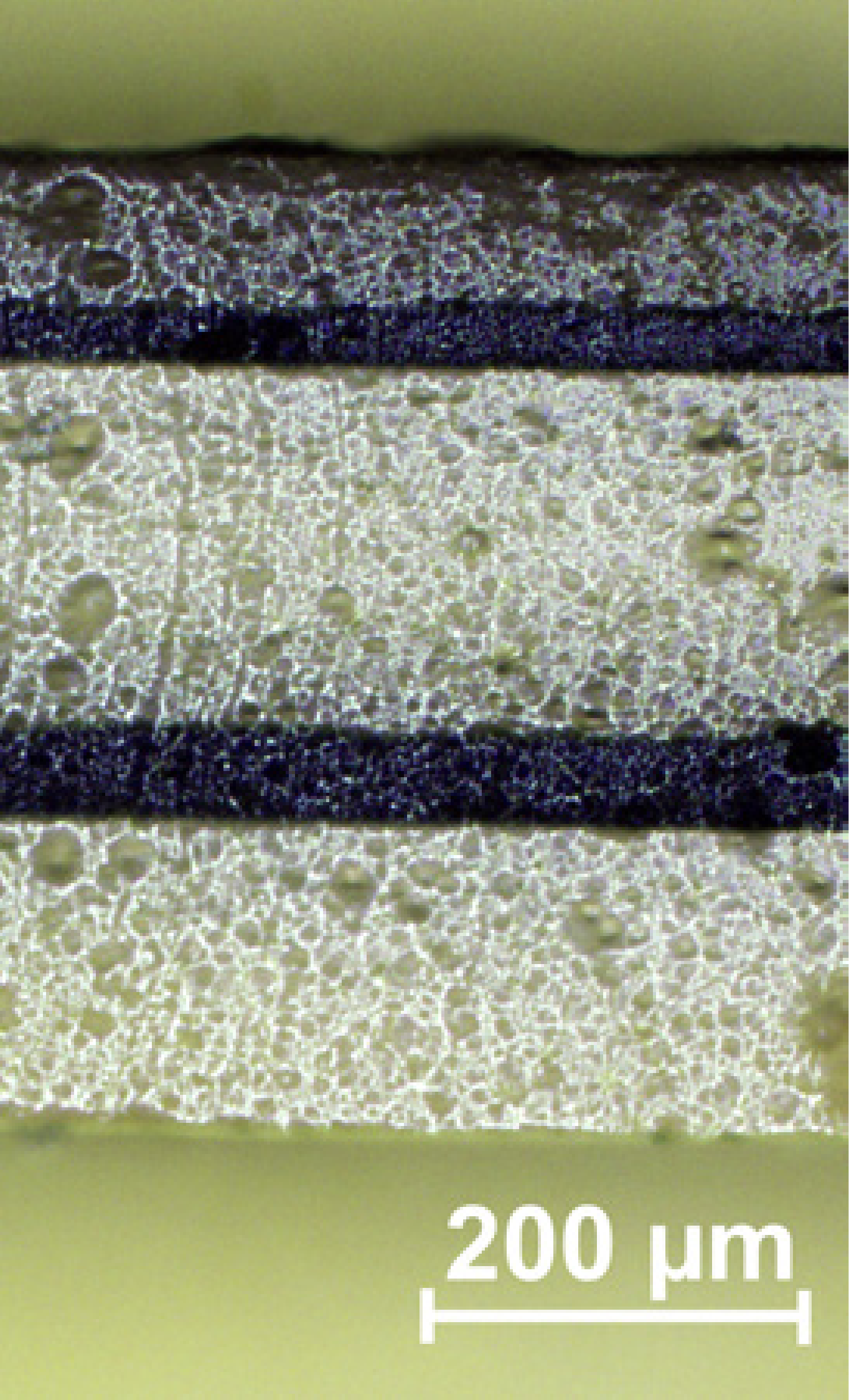}
\end{wrapfigure}
layers are \unit[45]{\micro m} thick each (for the basic protective layer we use 4 layers of offset tape, and for the dielectric layer 2 layers of offset tape). The inset on the right shows a cross section of the sensor layers under a microscope.
The sheet resistance of a conductive layer is in the order of \unit[1]{kOhm} (four-point probe). The stiffness (Young's Modulus) of the pure layered RTV is \unit[729.6$\pm$13.4]{kPA}, with two embedded conductive layers \unit[979.6$\pm$16.6]{kPA} (calculated from three samples each with the setup and method as described in \cite{Hopf2016}).

\begin{figure}
	\centering
	\includegraphics[width=1\linewidth]{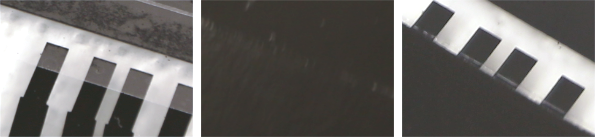}
	\caption{Left: sensor after casting the dielectric layer, the connector pads are covered by transparent sticky tape. Middle: after casting the second conductive layer. Right: after removal of the sticky tape (before curing in the oven); the connector pads stay exposed.}
	\label{fig:offsetconnectors}
\end{figure}

\subsubsection*{Connectors.}
The electrode strips must be connected to our electronic boards for measurement (see Appendix \ref{sec:measuring} for details). During fabrication we cover the connectors with sticky tape  before casting the remaining layers. The tape is removed before curing the corresponding layer, re-exposing the connectors, see \figref{fig:offsetconnectors}.

\subsubsection*{Finalization.}
The sensor is cut to the desired outline shape with the laser cutter. The resulting sensor is then pulled off the glass plate, and silicone adhesive can be optionally used to close the sensor to form, for example, a cylinder (\figref{fig:sensor-flat-cylinder}) to wrap a wrist or an elbow.


\section{Surface deformation reconstruction}
\label{sec:deformation}

Our sensor is equipped with simple rest state geometry, represented by a triangle mesh $\mathcal{S} = (\mathcal{V}, \mathcal{F}),$ where $\mathcal{V}$ is the set of 3D vertex positions and $\mathcal{F}$ is the connectivity (the set of faces). The connectivity $\mathcal{F}$ comes from meshing the electrode layout (\secref{sec:array}): we represent each sensor cell $S_j$ with a fan of triangles and mesh the overall layout using Delaunay triangulation using~\cite{Triangle}. We set the rest state geometry $\mathcal{V}^0$ to the canonical shape corresponding to the chosen topology: e.g., for the sensor in \figref{fig:sensor-flat-cylinder} (right), we use a circular cylinder of dimensions corresponding to the intrinsic size of the produced sensor. As the sensor is pulled onto a deforming object and capacitance changes are measured, the goal is to reconstruct the deformed geometry $\mathcal{V}(t)$ for each frame $t$, given the measured capacitances $C_j(t)$ of all sensor cells $S_j$.

Through the relation of capacitance to area (\equref{eqn:face2cap}), our sensor provides rich, localized area change measurements at interactive frame rates, but areas alone are not sufficient to define the shape of a general deforming surface in 3D, since area is an intrinsic quantity. 
We therefore pair these measurements with a data-driven geometric prior, acquired by simultaneously capturing the deformation of the object of interest using our sensor and an optical tracking system, and then training a regressor that maps the capacitance measurements to marker vertex positions.

\begin{figure}
	\centering
	\includegraphics[width=1\linewidth]{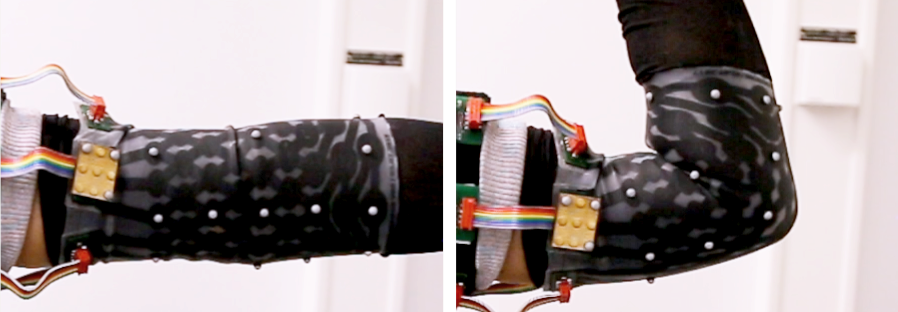}
	\caption{Our sensor on an elbow. Left: rest pose; right: close to fully bent.}
	\label{fig:sensorwithmarkers}
\end{figure}

To this end, we define a sparse set of vertex indices $\mathcal{M}$ and attach reflective markers onto the corresponding physical locations.
To simplify the marker attachment process, the set $\mathcal{M}$ is a subset of the mesh vertices corresponding to centers of circular sensor cells. The set is chosen to obtain a regular coverage of the cylindrical sensor, allowing a maximal distance of 5 centimeters in-between the individual markers. For all experiments we used a single, fixed marker pattern per sensor layout.
Placing the sensor onto the object of interest (\figref{fig:sensorwithmarkers}), we simultaneously record sensor readings and 3D marker positions tracked by an 8-camera OptiTrack setup \cite{Optitrack}.
Untreated silicone is highly specular, but we found that a matte finish can be attained by densely etching the outer layer on the laser cutter (with 60 Power, 100 Speed, and 500 Pulses/inch).
The captured and processed data for each frame $t$ consists of:
\begin{itemize}
\item Coordinate frame transformation $\mathbf{T}(t) \in \R^{3\times 4}$ (a $3\times 3$ rotation and a translation, recovered from 3 designated markers);
\item Marker positions $\mathbf{p}_i(t) \in \R^3$ w.r.t.\ the local frame, for each marker vertex $i \in \mathcal{M}$;
\item A vector $\mathbf{C}_c(t)$ of capacitance values of all sensor cells, obtained as described in \secref{sec:array}.
\end{itemize}
This data is used to train a regressor $g_{\theta}(\mathbf{C}_c(t))$ that maps sensor cell capacitance values to marker vertex position estimates $\hat{\mathbf{p}}$. Given $g$, we can employ the sensor at run-time and use the marker positions predicted by $g$ as positional constraints that guide the deformation of the sensor mesh $\mathcal{S}$.

\subsection{Capturing and processing training data}
\label{sec:mocap-cleaning}
\begin{figure}[t]
	\centering
	\includegraphics[width=1\linewidth]{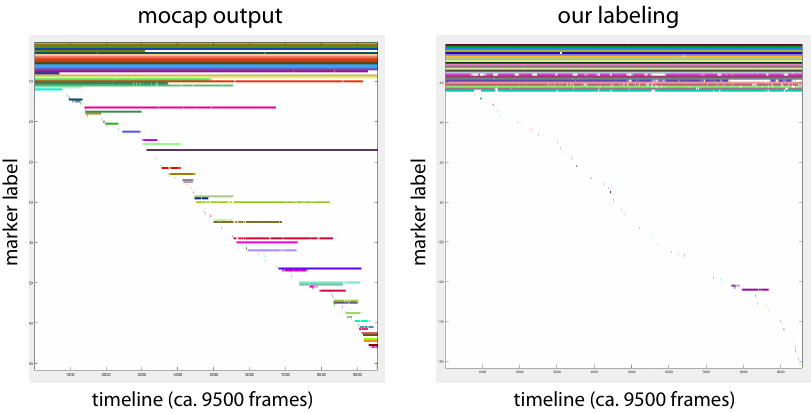}
	\caption{Marker labeling. For each individual label, we plot horizontal bars spanning the frames where it is visible.  Left: Captured markers directly from the mocap system. There are 165 individual labels due to periods of occlusion and subsequent failure to pick up the track, despite the actual number of markers being only 12. Right: sanitized and relabeled markers using our semiautomatic approach. A minority of outliers remain in a few frames; they are discarded from the dataset.}
	\label{fig:labeling}
\end{figure}

A fundamental challenge with marker based approaches are incorrectly labeled or lost markers, an issue exacerbated in settings like ours, where heavy occlusions and strong non-rigid deformations are combined with the lack of a simple skeletal prior. \figref{fig:labeling} provides an illustrative example of tracking 12 wrist-mounted physical markers. The OptiTrack system outputs 165 individual marker observations due to frequent tracking failures (sequence length is $1.5$ minutes). This problem quickly becomes unwieldy; in capturing real data we encountered more than 500 marker labels in a dataset of 17000 frames (3 minutes) of 21 physical markers.

\begin{figure}
	\centering
	\includegraphics[width=1\linewidth]{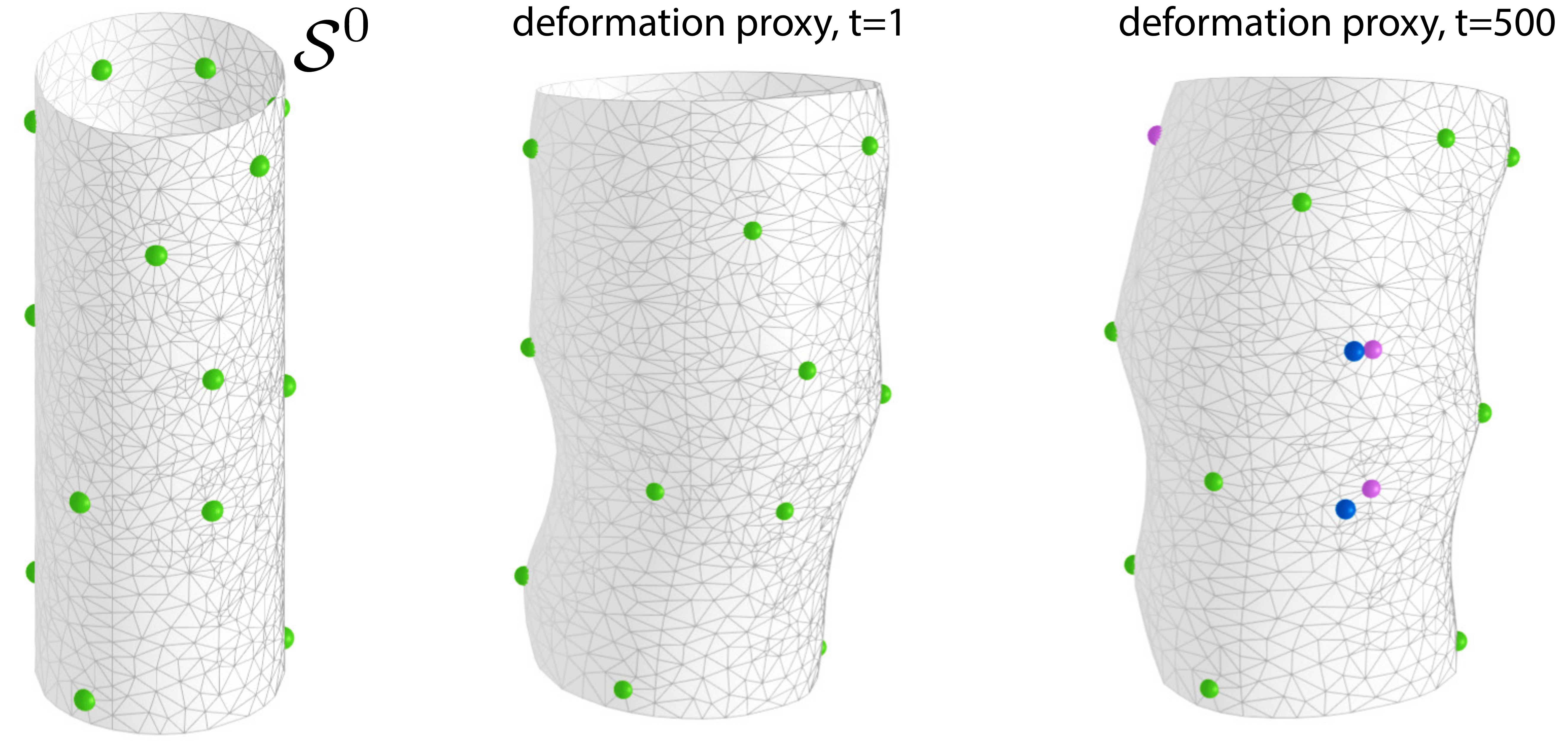}
	\caption{Left: The rest state sensor mesh $\mathcal{S}^0 = (\mathcal{V}^0, \mathcal{F})$ with marker vertices $\mathcal{M}$ in green. Middle: $\mathcal{S}^0$ deformed by the marker positions of the first frame in a wrist capture session. Right: the labeled markers in green, two unlabeled marker observations in blue and the two candidate matches in pink; the mesh geometry is estimated by elastically deforming $\mathcal{S}^0$ using the green markers as positional constraints.}
	\label{fig:labelestimation}
\end{figure}
Manual cleanup, label merging, and correct attribution would require hours of manual labor and make the acquisition of our deformation prior impractical. We therefore employ a novel semiautomatic marker cleanup and labeling pipeline.

The mocap system outputs a set of marker labels $\mathcal{I} = \{1, 2, \ldots, N\}$, and for each frame $t$, a binary indicator that tells whether the marker was visible in that frame. For each frame $t$ where marker $j$ is visible, the system also outputs its 3D position $\mathbf{p}_j(t) \in \R^3.$ 
We seek an assignment of marker vertices $i \in \mathcal{M}$ to tracked marker labels $j \in \mathcal{I}$, providing a 3D position in each frame $t$. Our main insight is to employ a state-of-the-art elastic deformation technique to create a proxy deformation of $\mathcal{S}^0$, using reliably labeled marker vertices as positional constraints. This allows us to match each unlabeled marker to its closest marker vertex on the proxy.

\paragraph{Initialization.} Usually the number of tracked labels $N$ is much larger than the actual number of physical markers, because some markers are temporarily lost, and are then given a new label when they re-enter. We initialize the assignment of marker vertex indices by picking $3$ tracked markers in the first frame and manually matching them with their corresponding mesh vertices in our rest pose mesh $\mathcal{S}^0$. We then rigidly transform $\mathcal{S}^0$ to align it with the tracked data (i.e., put it in the same coordinate system) by solving the Procrustes problem.

We then assign a 3D position to all remaining marker vertices of the mesh by searching for the closest tracked marker position in this frame. This way we obtain $|\mathcal{M}|$ pairings between marker labels and mesh vertex indices, as typically in the first frame (rest pose) all markers are visible. 

\paragraph{Labeling.}
We sort the unassigned tracked markers in chronological order according to the first frame they are visible at.

For each unassigned marker $j^*$ and for each frame $t$ where $j^*$ is visible, we elastically deform $S^0$ to match the captured geometry in $t$ by imposing the marker vertices in $\mathcal{M}$ that already have matched marker positions in frame $t$ as positional constraints. The output is a set of deformed ``proxy'' meshes, one for each such frame, which we use to find a match for $j^*$. For robustness, we pick the mesh vertex whose average $L^2$ distance over all frames is the smallest. We accept the match only if this distance is below a threshold $\tau$ (\unit[25]{mm} in our experiments), otherwise $j^*$ is marked as an outlier.

Every successful labeling provides an extra positional constraint for the deformations, improving the quality of the proxy (and thus the success rate) for subsequent labeling passes. In our implementation, we use the  deformation optimization method by Wang et al.~\shortcite{Wang15}, a state-of-the-art nonlinear elastic deformation technique that expects solely sparse positional constraints as input.

As a post-processing step, we visually inspect the produced assignments via 3D renderings and plots of $x,y,z$ coordinates over time, to detect  incorrect merges. If any are present, we can separate them and rerun the labeling algorithm again. One iteration of this procedure was sufficient for most of our capture sessions.

Our MATLAB implementation takes below 15 minutes per session, allowing us to have a 3 minutes long  captured session cleaned in around 10 minutes.
Note that we are not guaranteed to find observed 3D positions for each marker vertex of our mesh in each and every frame $t$, due to occlusions, outliers and possible failures of our assignment heuristic. We thus discard frames with unassigned markers, which are around 20 \% in our acquisition sessions. We encountered one case where too many markers were missing in some frames due to heavy occlusions in the folded elbow, which hampered the regressor training due to insufficient data. We resorted to synthetic 3D data for those frames, taking missing marker vertex positions from the deformation proxy.

\subsection{Regressor training}
\label{sec:regressor}
We wish to recover dense surface deformations in real time.

To this end, we learn a function $g_{\theta}(\mathbf{C}_c)$, parametrized by a deep neural network, that maps from sensor cell capacitances $\mathbf{C}_c \in \R^{s}$ to marker positions $\hat{\mathbf{p}} \in \R^{3\times|\mathcal{M}|}$ (in a local frame). We have experimentally verified that nonlinear function approximators such as the fully connected multi-layered neural network used here, perform better than linear models due to the nonlinearities in the mapping from area change to capacitance (Table \ref{table:lr-svm-nn-table}).

\begin{figure}
	\centering
	\includegraphics[width=0.8\linewidth]{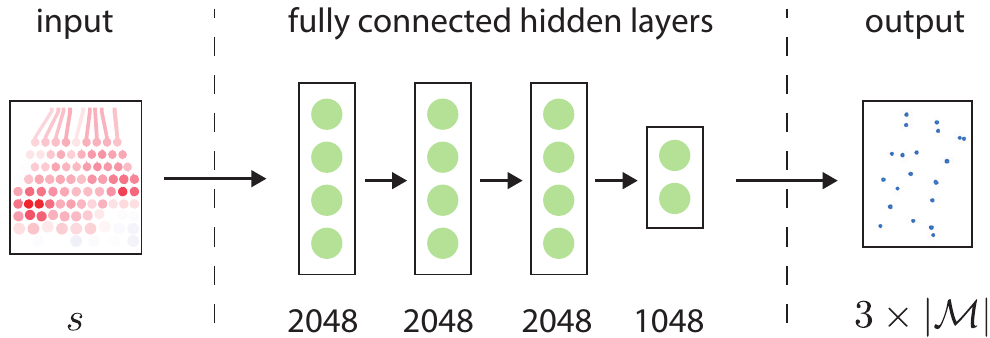}
	\caption{To train a sensor with $s$ sensor cells and $|\mathcal{M}|$ markers, our network takes $s$ capacitance readings as input and outputs $|\mathcal{M}|$ vertex position estimates, through three fully connected layers with 2048 units each and one fully connected layer with 1024 units. E.g. for our sensor in \figref{fig:sensor-flat-cylinder} there are 92 inputs and 63 ($3 \times 21$) outputs.}
	\label{fig:neuralnetwork}
\end{figure}

Our network architecture, depicted in \figref{fig:neuralnetwork}, takes $s$ sensor cell capacitance readings as inputs of a linear layer, followed by three fully connected layers with 2048 units each and one fully connected layer with 1024 units. A final linear output layer predicts the marker vertex positions $\hat{\mathbf{p}}$. The input and all hidden layers are followed by a ReLu activation function and a BatchNorm layer.
Given a training set $\mathcal{D}=\{(\mathbf{C}_c^i,\mathbf{p}^i)\}$ of $K$ vectorized ground truth input-output pairs, we perform
training via a weight-regularized $L^2$ loss:
\begin{align}\label{eq:loss-regression}
\mathcal{L}_\text{reg} = \sum_{i=1}^{K}
	\normltwo{g(\mathbf{C}_c^i) - {\mathbf{p}}^i} + \lambda \normltwo{\theta}
	\ ,
\end{align}
where $\theta$ are the model parameters and $\lambda$ is a regularization factor.

We implement the network using pyTorch \cite{pyTorch} and train it with the ADAM optimizer with a learning rate of $10^{-4}$, mini-batch size of $256$, regularization $\lambda = 10^{-5}$ and default values for all other parameters \cite{kingma2014adam}. All inputs are normalized to be zero-mean unit variance.

\subsection{Capturing dense surface deformation at runtime}

Once the neural network is trained and the regressor $g_{\theta}$ is available, we can deploy our sensor standalone, uncoupled from the optical tracking and estimate the dense surface deformation of an object without line-of-sight. This is illustrated in \figref{fig:no-line-of-sight-figure}, where the sensor is worn underneath clothing, rendering vision based approaches infeasible.
The regressor provides 3D positions $\mathbf{\hat{p}} = g_{\theta}(\mathbf{C}_c)$ of the marker vertices $\mathcal{M}$ given current sensor measurements $\mathbf{C}_c$. We note that the network is able to compensate for inaccuracies in area estimates  from capacitive readings (see \figref{fig:flatstretchtest}), which in particular occur under extreme stretch (see \secref{sec:results_accuracy}). To reconstruct the current surface deformation, we deform the rest state mesh $\mathcal{S}^0$ using the method proposed by Wang et al.~\shortcite{Wang15}, where the marker vertices $\hat{\mathbf{p}}$ again serve as positional constraints.

\section{Experiments and results}

\begin{figure}
	\centering
	\includegraphics[width=0.8\linewidth]{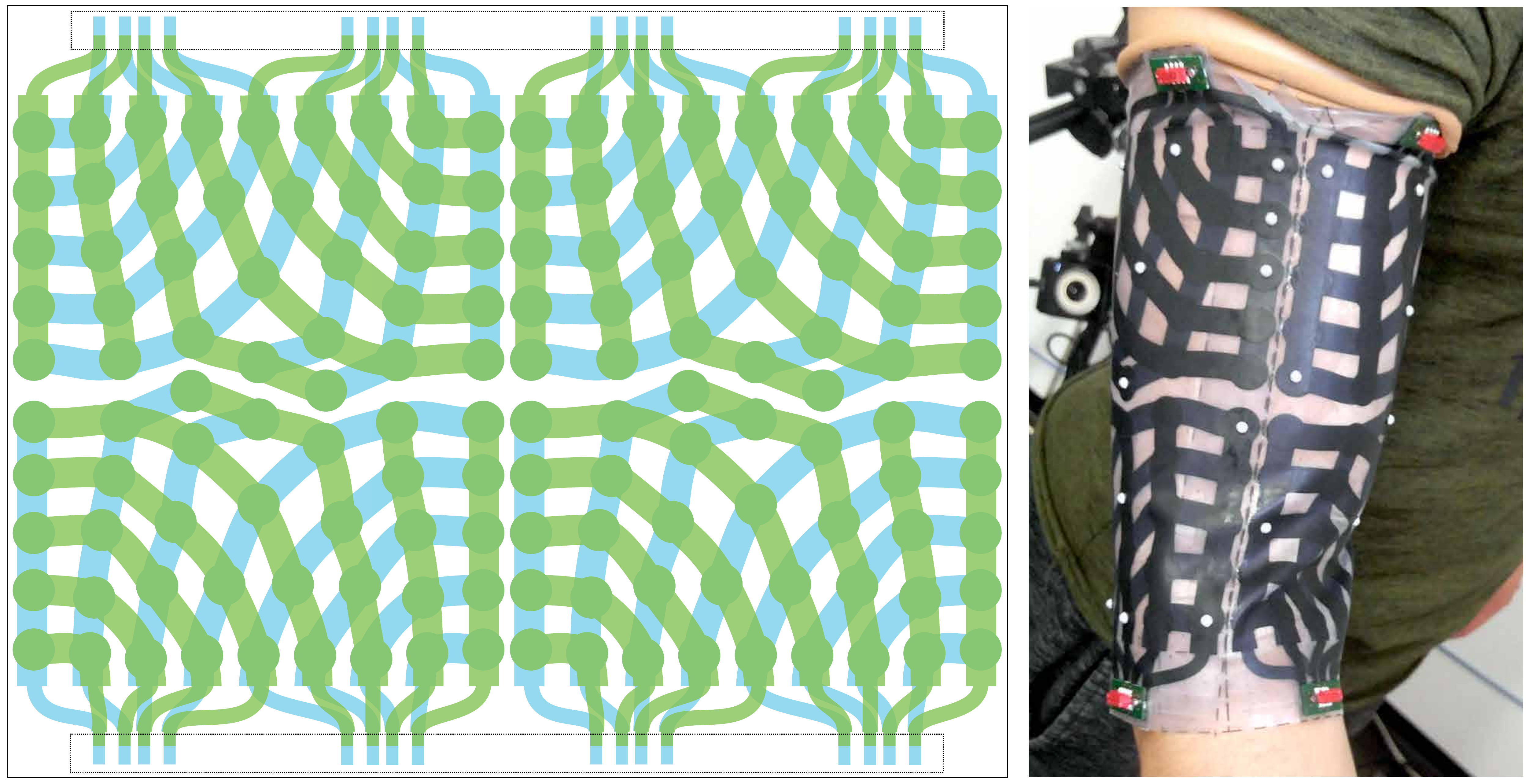}
	\caption{We fabricated a second larger sensor (300\textsf{x}250 mm) with 144 sensor cells and connectors on two sides. Left: The sensor layout consists of four identical sub-sensors that can be read out in parallel. Right: The produced sensor, glued to form a cylinder and worn on a biceps.}
	\label{fig:bicepslayoutphoto}
\end{figure}

To demonstrate the utility of our proposed approach, we evaluate its components in an ablative manner.
First, we quantitatively assess the sensor concept and the corresponding fabrication method (\secref{sec:experiments1}) and then demonstrate the applications in reconstruction of surface deformations, both qualitatively and quantitatively (\secref{sec:experiments2}).
Our experiments are performed with two  sensor layouts, shown in Figures \ref{fig:sensor-flat-cylinder} and \ref{fig:bicepslayoutphoto}. The layouts are manually designed, non-uniform grids, with all strips routed to the same side of the sensor, where they are connected to a connector PCB. The first layout is used both in its flat form and as a cylinder.

\subsection{Sensor characterization}
\label{sec:experiments1}

\subsubsection*{Distance sensor comparison.} 
\begin{figure}
	\centering
	\includegraphics[width=0.75\linewidth]{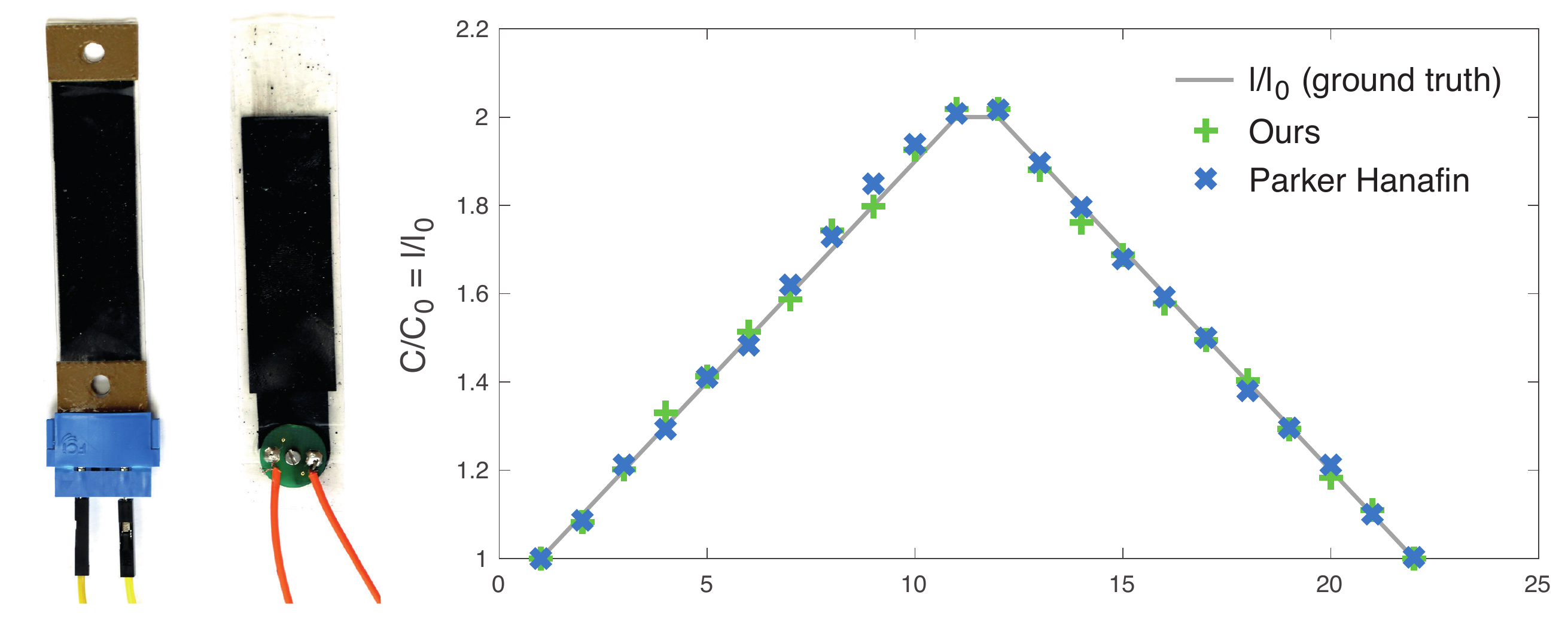}
	\caption{Left: An industrial sensor by Parker Hanafin and a sensor of the same dimension fabricated by us. Right: Comparison of their accuracy.}
	\label{fig:the1dtest}
\end{figure}

We verify the accuracy of our sensors by fabricating a uni-axial sensor with the same dimensions (15$\times$\unit[50]{mm}) as a commercially available Parker Hannifin industrial 
\setlength{\intextsep}{3pt}%
\setlength{\columnsep}{8pt}%
\begin{wrapfigure}{r}{0.5\linewidth} 
    \includegraphics[width=\linewidth]{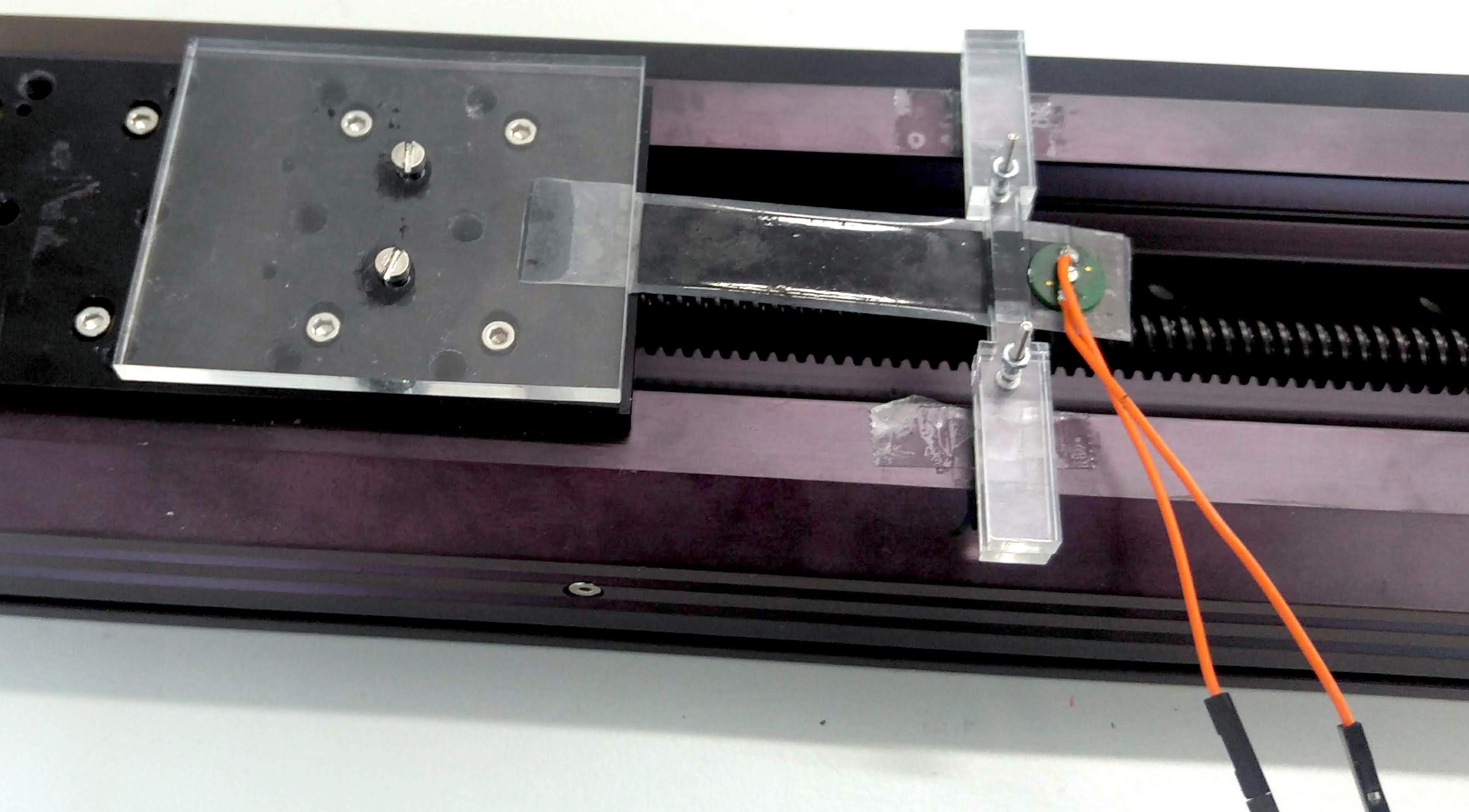}
\end{wrapfigure}
sensor \cite{ParkerHannifin}. We stretch both sensors (with a motorized linear stage, see inset) to various lengths and directly compare the
 readings. The average relative error of the two sensors is comparable (\figref{fig:the1dtest}), with a slight but non-significant edge for the Parker Hanafin sensor (0.0085) over ours (0.0096). Overall, we conclude that the accuracy of our measurements is high and comparable to commercial solutions. We note that there was no observable hysteresis in our experiments.

\subsubsection*{Longterm sensor behavior} 
\begin{figure}
	\centering
	\includegraphics[width=1\linewidth]{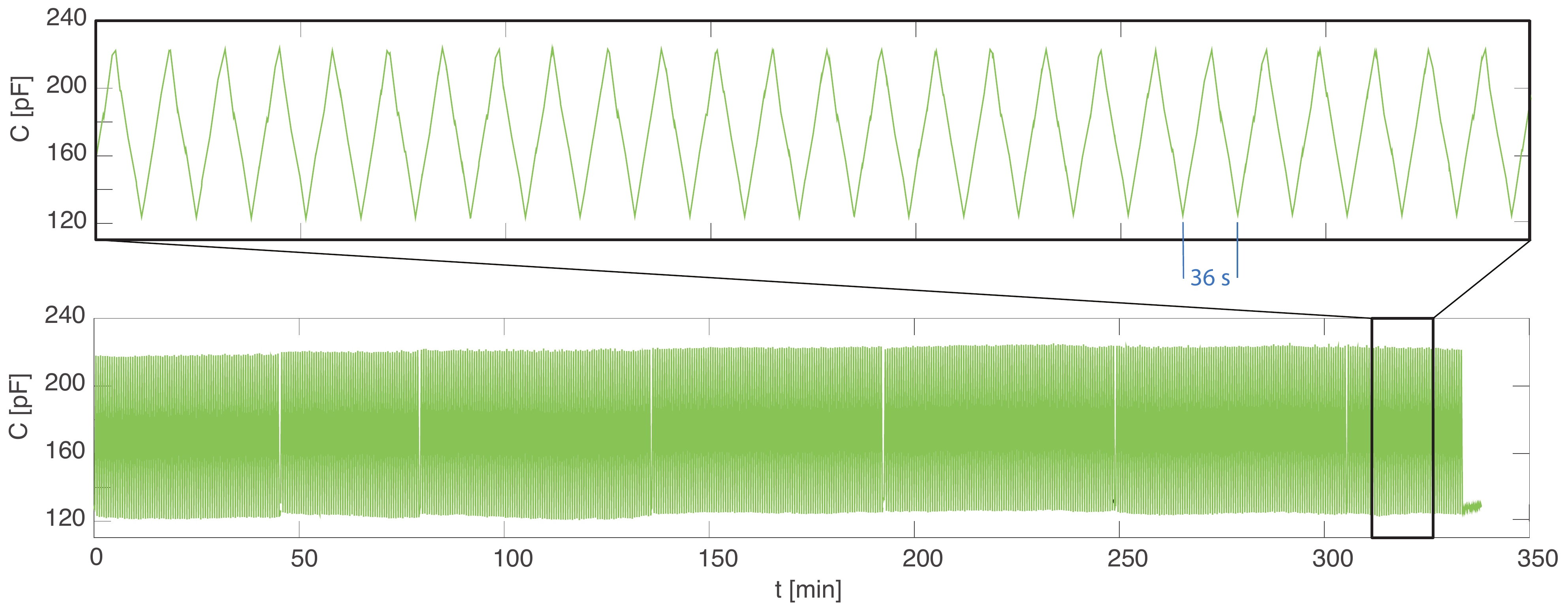}
	\caption{The uni-axial sensor response stays constant during a cyclic stretch ($2$\textsf{x}) test of 5 hours and 30 minutes (about 550 cycles).}
	\label{fig:longtermstretch}
\end{figure}

\begin{figure}
	\centering
	\includegraphics[width=0.95\linewidth]{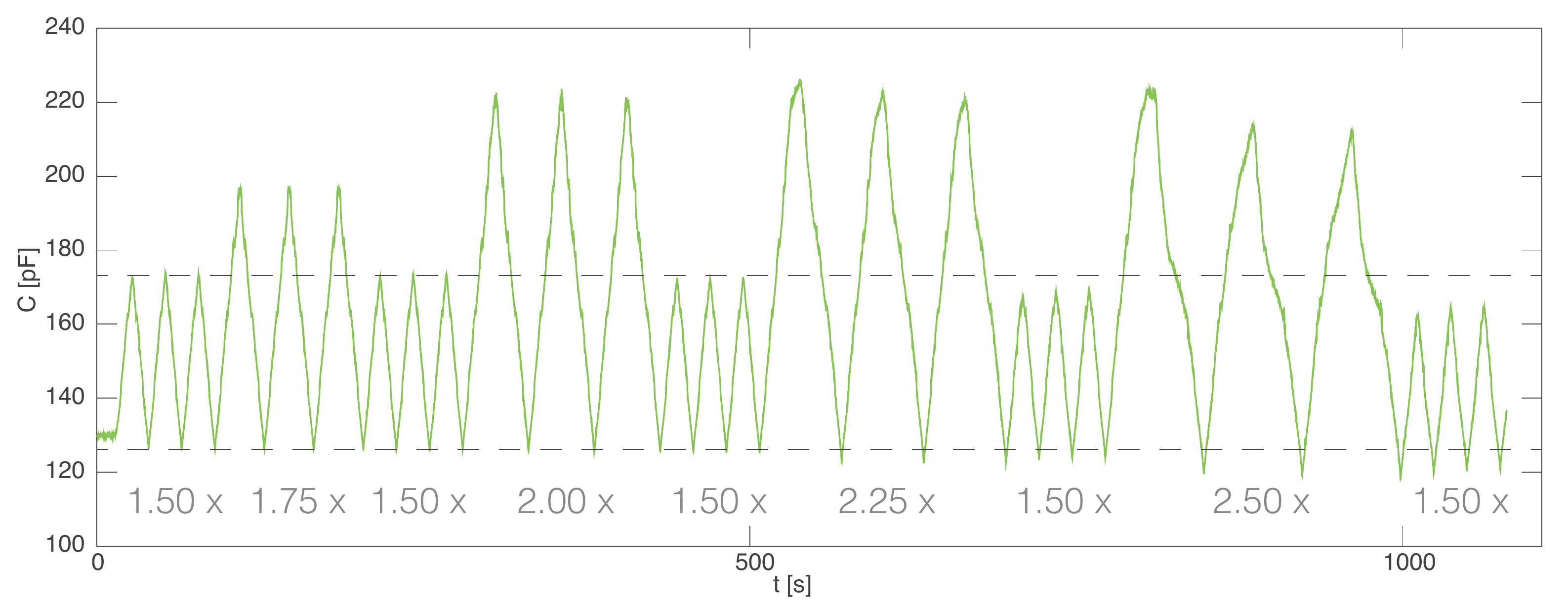}
	\caption{After a stretch factor of 2.25\textsf{x}, the sensor response when stretched by a factor of  1.5\textsf{x} has changed compared to the first three rounds.}
	\label{fig:maxstretch}
\end{figure}

In a second set of experiments, we evaluate whether and how the sensor response changes under longterm cyclic stretch and large stretch. For the longterm experiment, the uni-axial sensor is pre-stretched a few times and then continuously stretched and relaxed for \unit[5]{h} \unit[30]{min} by a factor of $2$\textsf{x}. The sensor response stays constant (see \figref{fig:longtermstretch}). The maximally allowed stretch before (internal) material damage occurs is found by stretching the sensor a few times to a baseline factor of $1.5$\textsf{x}, increasing the maximum stretch factor in each round (see \figref{fig:maxstretch}). These experiments show that our fabricated sensor can be stretched without noticeable internal damage by 100\% ($2$\textsf{x}) for at least \unit[5]{h} \unit[30]{min}. In our experiments, this stretch factor was never surpassed when capturing body parts.

\subsubsection*{2D stretch localization.}
To assess the localization capabilities of our sensor layout, we perform a simple experiment, in which we fix a flat sensor to a frame and poke it in different locations. \equref{eqn:area_to_capacitance} states that the sensor cells' capacitance changes directly relate to area changes. The proposed readout scheme (cf.\ \secref{sec:readout}) allows us to measure and localize stretch. \figref{fig:poketest} visualizes two example frames extracted from the video in the supplemental material. This capability could be explored in other application scenarios, including detection of touch and pressure.

\begin{figure}
	\centering
	\includegraphics[width=0.8\linewidth]{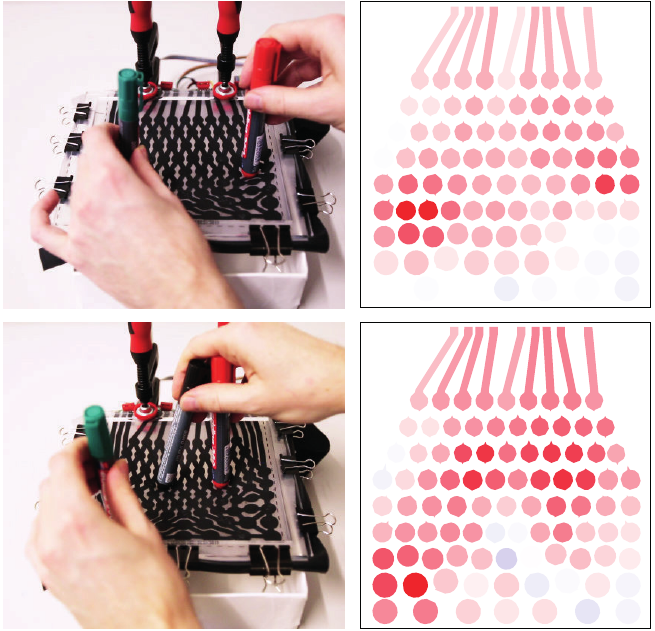}
	\caption{\label{fig:poketest}Left: the sensor is fixed to a frame and poked with pens. Right: Area change magnitude measured per sensor cell.
	}
\end{figure}

\subsubsection*{2D stretch quantification.}
\begin{figure}
	\centering
	\includegraphics[width=1\linewidth]{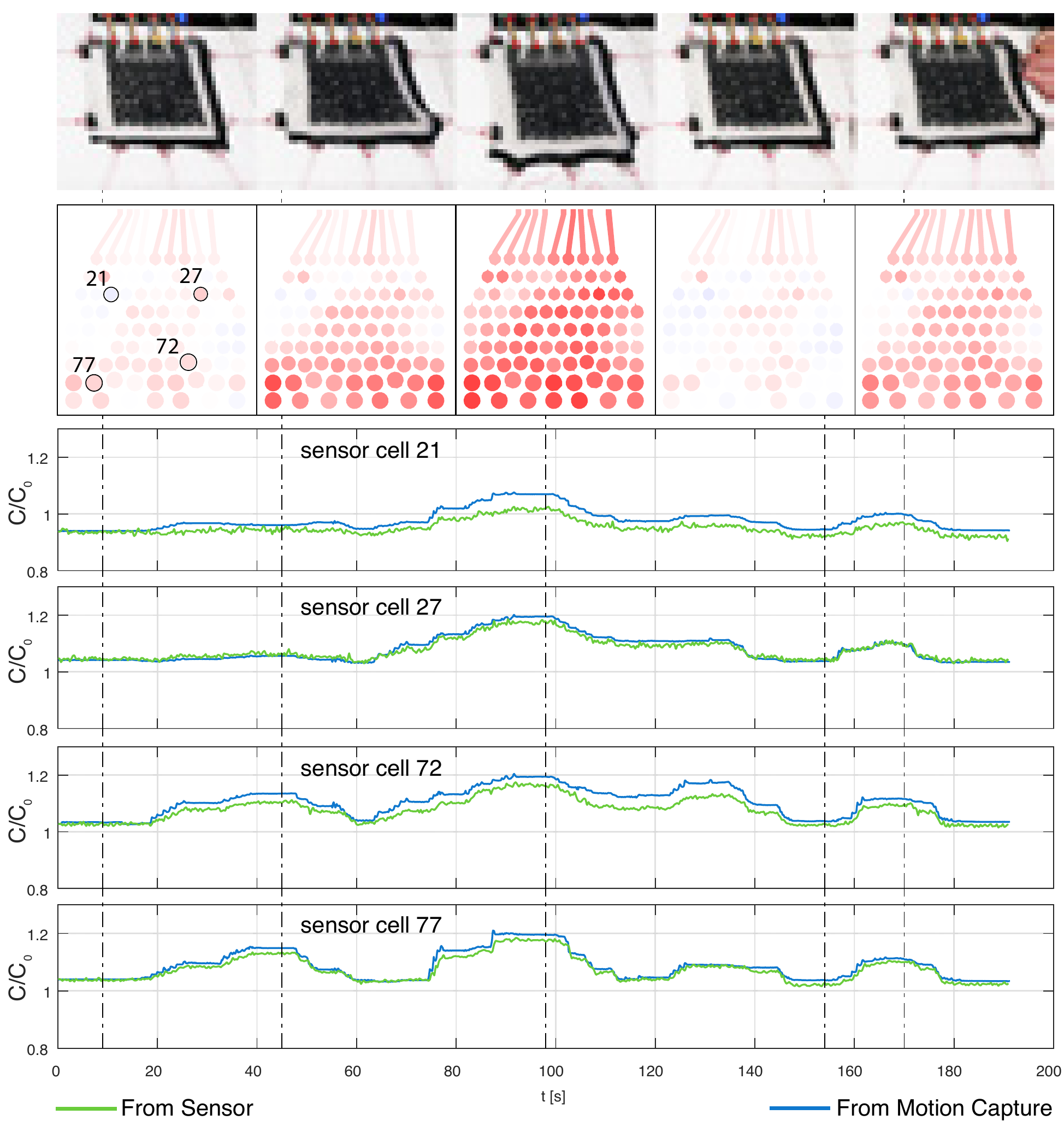}
	\caption{The sensor is dynamically stretched by selectively pulling on the strings its attached to. Top: A set of sample frames. Middle: Stretch intensity per cell at the sample frames. Bottom: The relative capacitance of four selected sensor cells over time, comparing ground truth (estimated through mocap markers) in blue and the capacitance change recorded by our sensor in green. The dashed vertical lines show the locations of the sample frames on the timeline.}
	\label{fig:flatstretchtest}
\end{figure}

To better understand the accuracy of recovered stretch measurements, we attach clips on strings to a flat sensor, so that we can apply spatially varying tension forces by selectively pulling on the strings. Additionally, we place reflective markers on the sensor, so that we can estimate the actual stretch per sensor cell.  \figref{fig:flatstretchtest} visualizes the results. We report an average relative error of 7.7\% when comparing the measured capacitance ratio $C_c/C_c^0$ with the theoretical capacitance ratio calculated by  \equref{eqn:face2cap} per sensor cell from the tracked areas. This error is likely due to our approximate sensor model, which neglects the influence of the (changing) resistance of the electrodes. Close inspection of \figref{fig:flatstretchtest} reveals that this effect is negligible for our purposes.

\subsection{Surface deformation capture}
\label{sec:experiments2}
\label{sec:results_accuracy}

\subsubsection*{Predictor comparison.} To validate our design choice of parameterizing the regression problem of \equref{eq:loss-regression} with a neural network, we perform a comparison with several alternative models as baseline. \tableref{table:lr-svm-nn-table} summarizes the results of a three-way comparison with linear regression and non-linear SVM using an RBF kernel. The neural network achieves the lowest mean and max errors and produces the lowest standard deviation across all datasets used in our experiments.

\begin{table}
\centering
\begin{threeparttable}
	\caption{Comparison of prediction accuracy of the chosen DNN regressor (\emph{ours}) with a linear regression model (LR) and a non-linear support vector machine with an RBF kernel (SVM). All errors are in millimeters, lower is better.}
{%
\small
\begin{tabular*}{\linewidth}{ll @{\extracolsep{\fill}} rrr}
\toprule
Marker error & & mean & std & max \\ \midrule

Balloon & LR & 3.59 & 1.90 & 12.84 \\
	  & SVM & 3.22 & 2.73 & 25.05 \\
      & \emph{ours} & \textbf{2.75} & \textbf{1.86} &  \textbf{12.85} \\ \midrule

Biceps & LR & 7.64 & 5.06 & 53.00\\
	  & SVM & 6.86 & 5.18 & 52.24 \\
      & \emph{ours} & \textbf{3.85} & \textbf{2.39} & \textbf{25.81}  \\ \midrule

Elbow & LR & 7.65 & 3.31 & 39.95 \\
      & SVM & 6.73 & 5.26 & 59.79 \\
      & \emph{ours} & \textbf{3.46} & \textbf{2.48} & \textbf{30.82} \\ \midrule

Wrist & LR & 12.8 & 4.99 & 71.89\\
	  & SVM & 4.36 & 2.71 & 44.12 \\
      & \emph{ours} & \textbf{3.51} & \textbf{2.14} & \textbf{27.22}  \\ \midrule
      
Forearm & LR & 10.64 & 3.94 & 52.03\\
	  & SVM & 4.38 & 2.30 & 32.11 \\
      & \emph{ours} & \textbf{4.02} & \textbf{2.66} & \textbf{38.52}  \\ \bottomrule

\end{tabular*}
}
\label{table:lr-svm-nn-table}
\end{threeparttable}
\end{table}

\subsubsection*{Non-skeletal 3D deformation.} 
\begin{figure}
	\centering
	\includegraphics[width=1\linewidth]{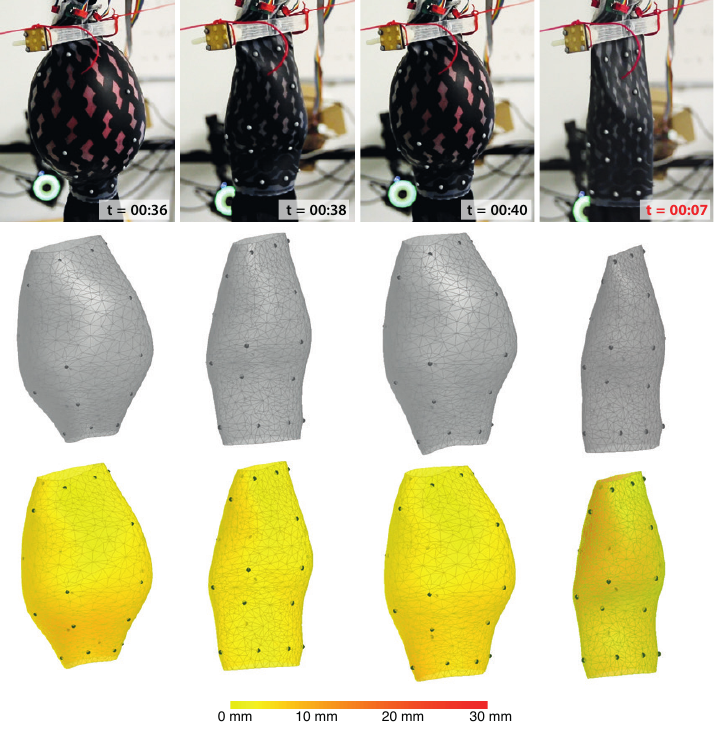}
	\caption{Four frames of a 1:40 min long balloon capture session. Top: Video frames for comparison. Middle: Mocap ground truth. Bottom: Reconstruction based on the sensor measurements and the trained prior. The rightmost frame corresponds to the frame with the largest individual marker error.}
	\label{fig:balloon}
\end{figure}

To demonstrate the deformation capture abilities of our sensor, we use it to measure the shape of a balloon that is aperiodically inflated (up to a maximum diameter of about 120 mm) and deflated. Despite the apparent simplicity of the setup, the deformation is freeform, and it is not possible to rely on standard geometric priors, such as a skeleton. We captured a 5-minute session with the mocap system (2451 frames), and used the cleaned data to train a regressor (\secref{sec:regressor}). To validate the system, we recorded an additional 1:40 min sequence (946 frames). The errors between our regressor and the mocap output are small, \unit[2.75]{mm} on average, with a maximum of \unit[12.85]{mm} (\figref{fig:balloon}, rightmost column). Note that the maximal resolution of our mocap system, which is used as ground truth for these measurements, is  \unit[0.2]{mm}.  \figref{fig:balloon} shows four frames extracted from the video in the supplemental material.

\begin{figure}
	\centering
	\includegraphics[width=1\linewidth]{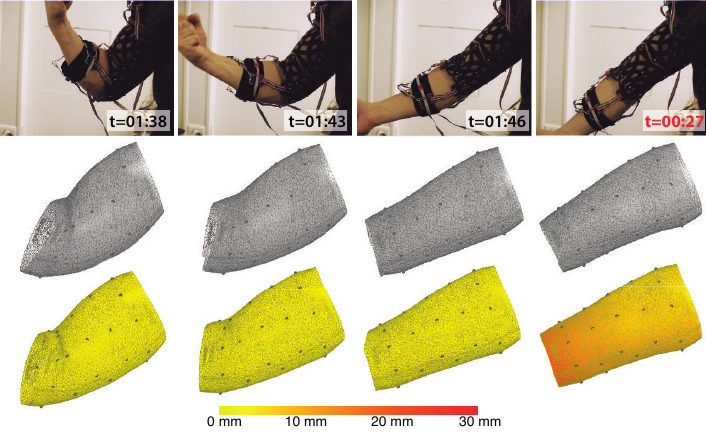}
	\caption{Four frames of a 2-minute long biceps capture session. Top: Video frames for comparison. Middle: Mocap ground truth. Bottom: Reconstruction based on the sensor measurements and the trained prior. The rightmost frame corresponds to the frame with the largest individual marker error.}
	\label{fig:biceps}
\end{figure}

As a non-skeletal body part example, we captured a biceps muscle of ca.\ \unit[36]{cm} in circumference being flexed, together with a small part of the elbow, using a larger sensor (see \figref{fig:bicepslayoutphoto}). We captured a 6-minute training session with the mocap system (2305 frames) and an additional \unit[2]{min} test sequence (1224 frames). We report an average marker error of \unit[3.85]{mm}, with a maximum of \unit[25.81]{mm} (\figref{fig:biceps}, rightmost column). \figref{fig:biceps} shows four frames (extracted from the video in the supplemental material).

\subsubsection*{Uni-axial deformation.} 
\begin{figure}
	\centering
	\includegraphics[width=1\linewidth]{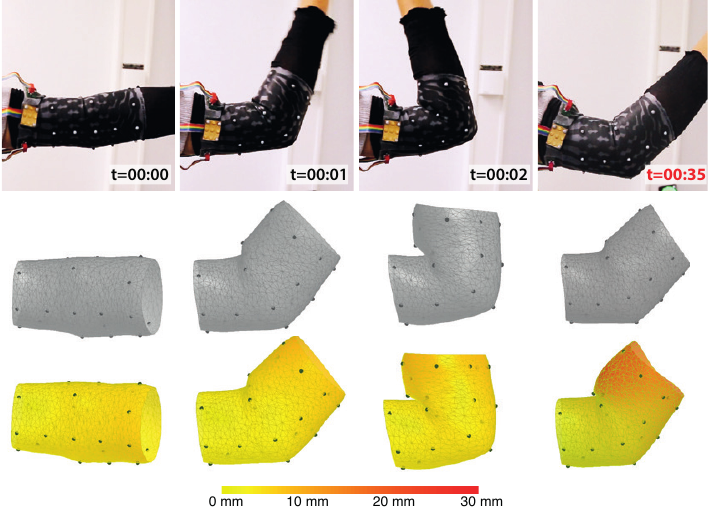}
	\caption{Four frames of an elbow capture session. Top: Video frames for comparison. Middle: Mocap ground truth. Bottom: Reconstruction based on the sensor measurements and the trained prior. The rightmost frame corresponds to the largest individual marker error.}
	\label{fig:elbow}
\end{figure}

We wrap our sensor around an elbow to capture its movement. This is a challenging scenario due to the strong occlusions when the elbow is fully bent and due to the local non-rigid surface deformation. We use 12 minutes of training data (5369 frames) and a 2-minute test sequence (1329 frames). Our sensor accurately matches the test sequence (\figref{fig:elbow}) and enables  deformation sensing even when worn below clothing (\figref{fig:no-line-of-sight-figure}). In this example, the mean error is \unit[3.46]{mm} and max error is \unit[30.82]{mm}. In \figref{fig:elbow} we show four frames extracted from the full video sequence (attached in the supplemental material).

\subsubsection*{Multi-axial deformation.} 
\begin{figure}
	\centering
	\includegraphics[width=1\linewidth]{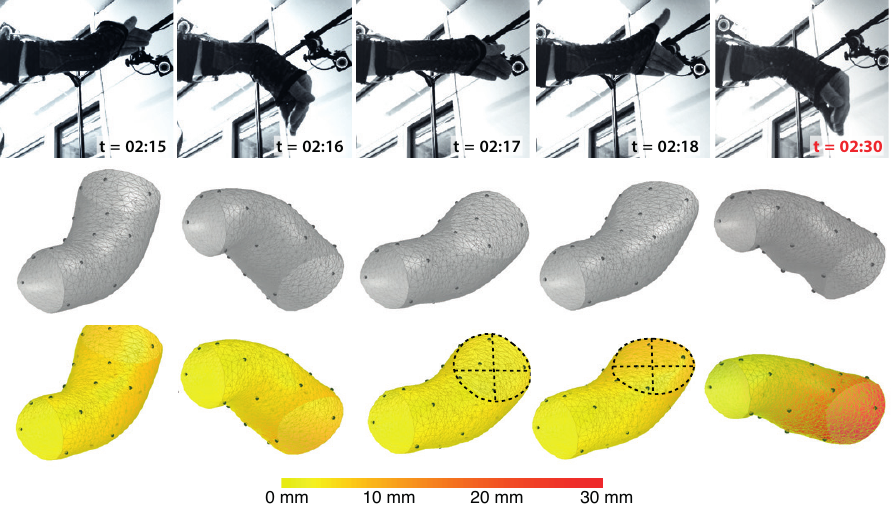}
	\caption{Four frames from a wrist capture session. Top row:  Video frames for comparison. Middle: Mocap ground truth. Bottom: Reconstruction based on the sensor measurements and  our trained prior. In the third and fourth frames, note how our sensor correctly senses its shape when the fingers are splayed. The frame corresponding to the largest individual marker error is shown on the right.}
	\label{fig:wrist}
\end{figure}

Our sensor successfully reconstructs very challenging scenarios, such as a wrist movement containing both a multi-axial skeletal deformation and volume changes when the fingers are splayed. For the wrist example, we trained on a 15-minute session (8799 frames), and tested on a 2:45 minutes session (1774 frames). Even in this case, the errors are low, with a mean of \unit[3.51]{mm} and max error of \unit[27.22]{mm} (see \figref{fig:wrist}).

\subsubsection*{Twisting motions.} 
\begin{figure}
	\centering
	\includegraphics[width=1\linewidth]{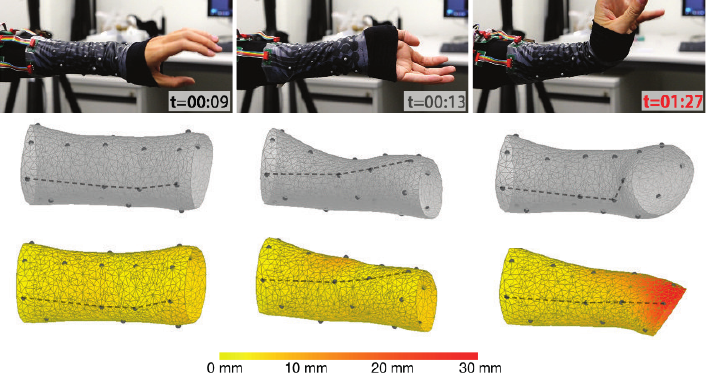}
	\caption{Three frames from the forearm capture session. Top row:  Video frames for comparison. Middle: Mocap ground truth. Bottom: Reconstruction based on the sensor measurements and  our trained prior. Be aware that our sensor is only able to capture local stretch occurring below the sensor. The frame with the highest individual error is shown on the right: The sensor fails to correctly predict the bending of the wrist.}
	\label{fig:twist}
\end{figure}

The sensor also manages to capture the twisting motion of a forearm. For this example the model is trained on a 8-minute session (1846 frames), and evaluated on a 2 minutes session (1320 frames). For such a scenario the errors are slightly higher with a mean error of \unit[4.02]{mm} and max error of \unit[38.53]{mm}, (see \figref{fig:twist}). The peak in error corresponds to predictions of the markers on the hand when the wrist is fully bent, see \figref{fig:twist} on the left.

\subsubsection*{Interpolation behavior} To demonstrate the robustness of our predictor in test situations with strains deviating from the training data, we artificially reduce the training data of the wrist example, while keeping the test set fixed. We only keep training frames where the angle $\alpha$ between the arm and the palm is $\alpha < \gamma$ or $\alpha > \beta$ ($\alpha$ is the angle between a line connecting two markers on the arm and another line connecting two markers on the back of the hand). \tableref{table:gen-table} shows the remaining number of training frames and the resulting mean and maximum error for a selection of angular limits. The first block (where frames with large angles are removed) shows that the network does not extrapolate well. Note that this is to be expected, since most machine learning approaches do not generalize well to situations where the training and test data statistics differ significantly. However, as shown in the middle and the lower block, the method manages to interpolate well, even though there are now training samples at shallow angles. This holds true as long as the training set is large enough. The last row of \tableref{table:gen-table} shows the results of exceeding this lower limit in terms of training data size.

\begin{table}
\centering
\begin{threeparttable}
	\caption{Predictor accuracy of the wrist test example with artificially reduced training data. It shows the ability of handling strains in the test data not previously seen during training. The training is reduced to frames with $\alpha < \gamma$ or $\alpha > \beta$, where $\alpha$ is the angle between the arm and the palm and $\gamma$, $\beta$ are angular limits.}
{%
\small

\begin{tabular*}{\linewidth}{rr @{\extracolsep{\fill}} rrr}
\toprule

$\gamma$ & $\beta$ & \#frames & mean & max \\ \midrule

$\alpha_\text{min}$  & $\alpha_\text{max}$ & 8799 & 3.51 & 27.22 \\ \midrule

60 & $\alpha_\text{max}$ & 8668 & 3.14 & 28.40 \\ 
40 & $\alpha_\text{max}$ & 7335 & 3.40 & 49.20 \\ 
30 & $\alpha_\text{max}$ & 5777 & 4.07 & 50.55 \\
20 & $\alpha_\text{max}$ & 3229 & 6.59 & 76.96 \\ \midrule

20 & 30 & 6251 & 3.35 & 26.45 \\
20 & 40 & 4693 & 3.89 & 31.31 \\ \midrule

$\alpha_\text{min}$ & 20 & 5570 & 3.41 & 35.76 \\
$\alpha_\text{min}$ & 30 & 3022 & 4.67 & 47.76 \\
$\alpha_\text{min}$ & 40 & 1464 & 7.38 & 52.50 \\ \bottomrule

\end{tabular*}
}

\label{table:gen-table}
\end{threeparttable}
\end{table}

\subsubsection*{Real-time reconstruction.} 
\begin{figure}
	\centering
	\includegraphics[width=1\linewidth]{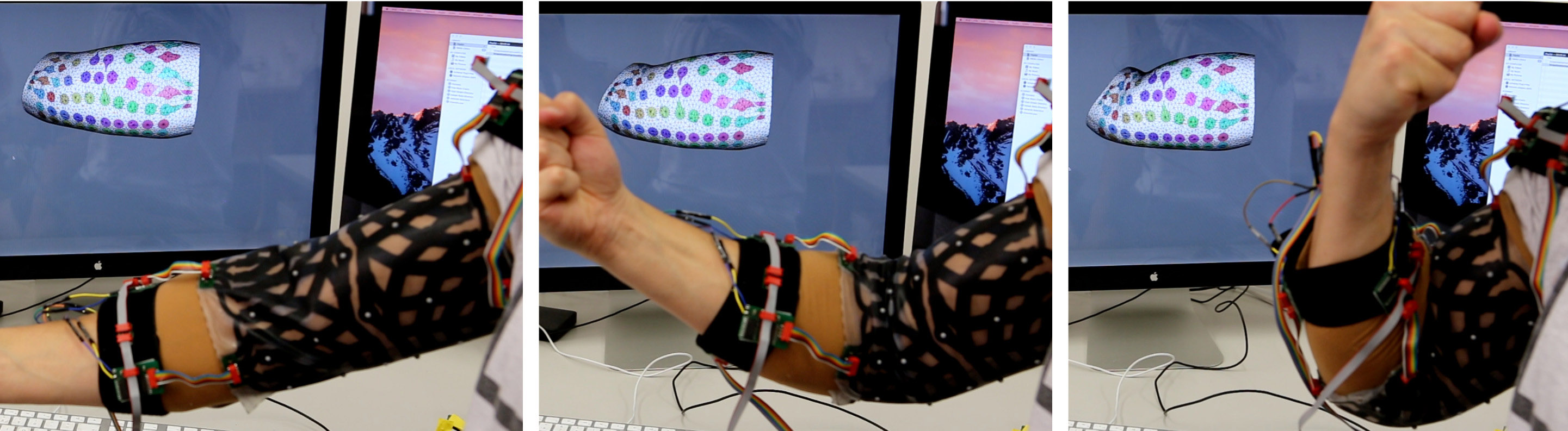}
	\caption{Three frames from a live capturing session of the biceps.}
	\label{fig:bicepslive}
\end{figure}

To demonstrate the real-time capabilities of our approach, we have implemented a live system in which a user may wear the sensor, and we deform a cylindrical (in rest pose) mesh at interactive rates (approximately \unit[8]{Hz}). See Figures \ref{fig:teaser},  \ref{fig:bicepslive}, and the accompanying video for the results. Note that in this setting, the users wear the sensor long after the training data was acquired; when taking the sensor off and putting it on again, one only needs to make sure that the alignment of the sensor and the body part is approximately the same. For the wrist example we quantitatively evaluated this effect of taking the sensor off and putting it on again with an imperfect alignment. For a 2-minute test sequence, the in-session mean error is \unit[4.06]{mm} (max: \unit[38.28]{mm}) while the out-of-session mean error is \unit[6.80]{mm} (max: \unit[47.22]{mm}).

\section{Limitations and Concluding Remarks}
We proposed a soft and stretchable capacitive sensor array that allows measuring localized area changes. When paired with a learned geometric prior, it can reconstruct complex deformations  without line-of-sight.

Our fabrication method and sensor layout open the door to multiple exciting future work venues. The most obvious is combining our area sensor with bend sensors to measure both extrinsic and intrinsic surface geometry, to e.g., also capture isometries. Furthermore it would be compelling to find a way to capture distance changes in such a dense array setting. These extensions would allow to estimate the deformation of general surfaces (like clothing) even if there's no non-area preserving stretching or twisting occurring. Another practical addition would be an assisting mechanism for correct placement of the sensor on the measured object: at present, we simply take a photograph before the training session and peruse it when putting the sensor on again for live session capture.  

The acquisition of a large dataset of training sequences with multiple users is necessary to generalize our approach to multiple users, skipping the per-user training session. As with other sensing modalities (e.g., EMG, EEG), additional research into solving the cross-session problem may be required in this setting. Furthermore, the computational design of sensor layouts that are optimized for a specific set of deformations is also an interesting challenge that would directly benefit from the flexibility and simplicity of our fabrication pipeline. Finally, more complex sensor (3D) geometries such as data gloves appointed with our sensor array would enable a number of compelling use cases, such as reconstructing fine-grained hand shape in real-time, sidestepping the various issues (occlusions, lighting) associated with other sensing modalities. 

We note that we employ a sparse set of markers as our \emph{ground truth}, and effectively reconstruct this set from our sensor readings. Ideally we would like to have densely captured 3D geometry for training, and match it to denser sensor readings. As discussed in \secref{sec:related}, spatially and temporally dense 3D capture is highly challenging and currently invariably involves some degree of model fitting. A realistic simulator that generates large quantities of high-quality synthetic data could be an alternative. It would be interesting to develop a denser version of our sensor design for more direct, dense geometry measurements. This comes with its own challenges, such as properly housing the electronic boards and a time multiplexing strategy to keep the read-out frame rates interactive; we leave this as future work.

\section*{ACKNOWLEDGMENT}

We would like to thank Denis Butscher, Christine de St. Aubin, Raoul Hopf, Manuel Kaufmann, Roi Poranne, Samuel Rosset, Michael Rabinovich, Herbert Shea, Rafael Wampfler, Yifan Wang, Wilhelm Woigk, Shihao Wu and Ji Xiabon for their assistance in the fabrication, with the experiments and for insightful discussions and Seonwook Park, Velko Vechev and Katja Wolff for their help with the video. This work was supported in part by the SNF grant 200021\textunderscore 162958, the NSF CAREER award IIS-1652515, the NSF grant OAC:1835712, and a gift from Adobe.

\bibliographystyle{ACM-Reference-Format}
\bibliography{bibliography}

\appendix

\section{Silicone Mixtures}
\label{sec:mixtures}

We used the following mixtures for the three types of silicone layers:

\noindent\textbf{Protective layer:} Silbione RTV 4420 \cite{Silbione} component A (weight ratio=1.0) and Toluol (1.0) are mixed, then Silbione RTV 4420 (1.0) component B is added.

\noindent\textbf{Conductive layer:} Silbione RTV 4420 component A (1.0) and Toluol (2.0) are mixed, then Silbione RTV 4420 (1.0) component B is added. In a separate container, Imerys Enasco 250 P \cite{Ensaco250} conductive carbon black (0.2) is mixed with isopropyl alcohol (2.0) by slowly adding the isopropyl alcohol while stirring. Then both compositions are combined and mixed for about 3 minutes. The 2-component silicone Silbione RTV 4420 was chosen due to its tear behavior as evaluated in \cite{Bernardi17} and the Imerys Enasco 250 P carbon black as suggested in \cite{Brunne2011}.

\noindent\textbf{Dielectric layer:} Same as the protective layer.

\section{Measurement setup}

\begin{figure}
	\centering
	\includegraphics[width=1\linewidth]{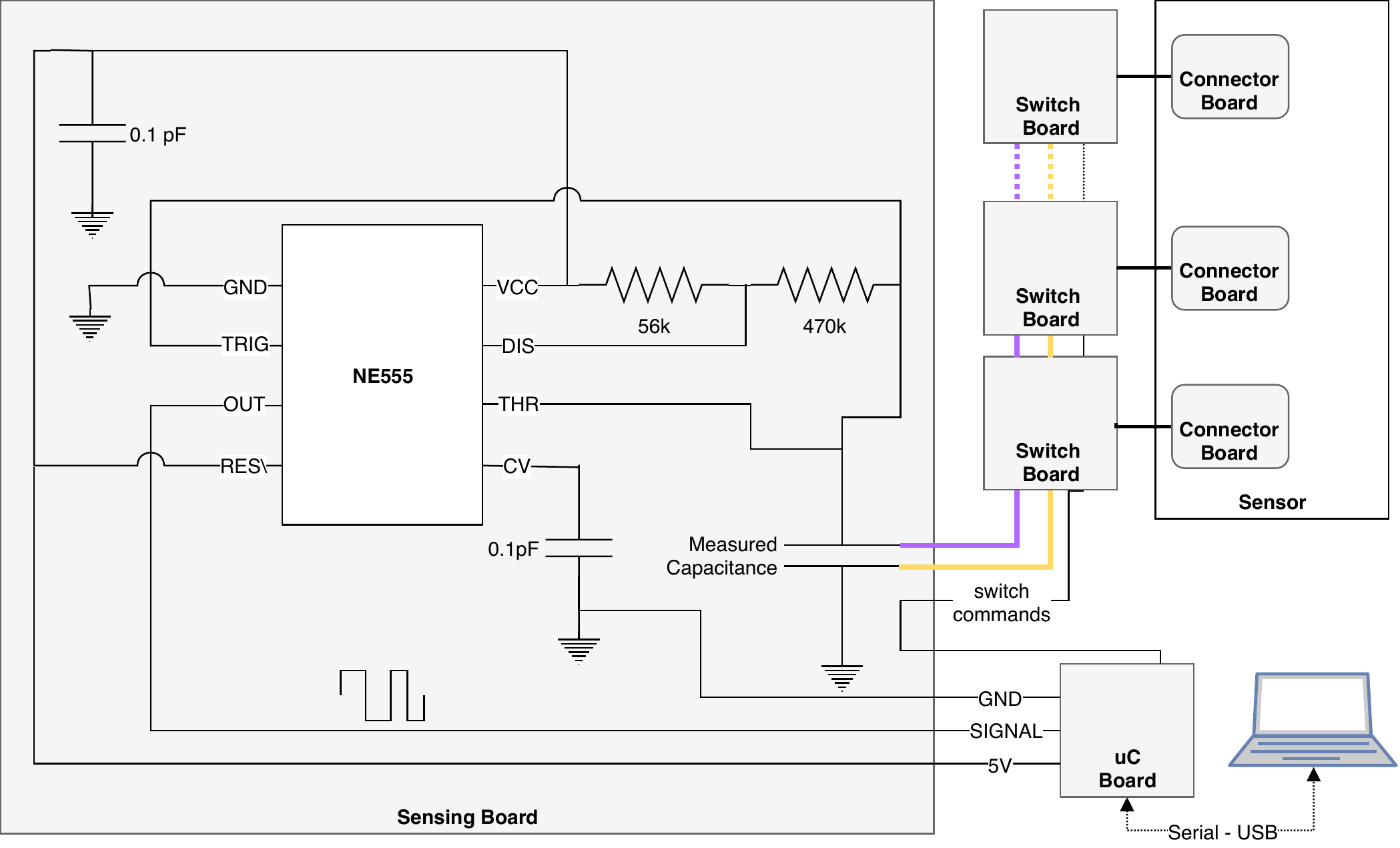}
	\caption{Our modular setup consists of two parts. Left: The capacitance sensing circuit is implemented with a NE555 timer IC, resulting in a square SIGNAL of the charging time that is read by the uC and sent to the computer. Right: The uC board and the switch boards go through all combinations, dynamically connecting the current set of source electrode strips (purple) and ground electrode strips (yellow); see \secref{sec:array} and \figref{fig:example3x3} for details.}
	\label{fig:circuit}
\end{figure}

\begin{figure}
	\centering
	\includegraphics[width=1\linewidth]{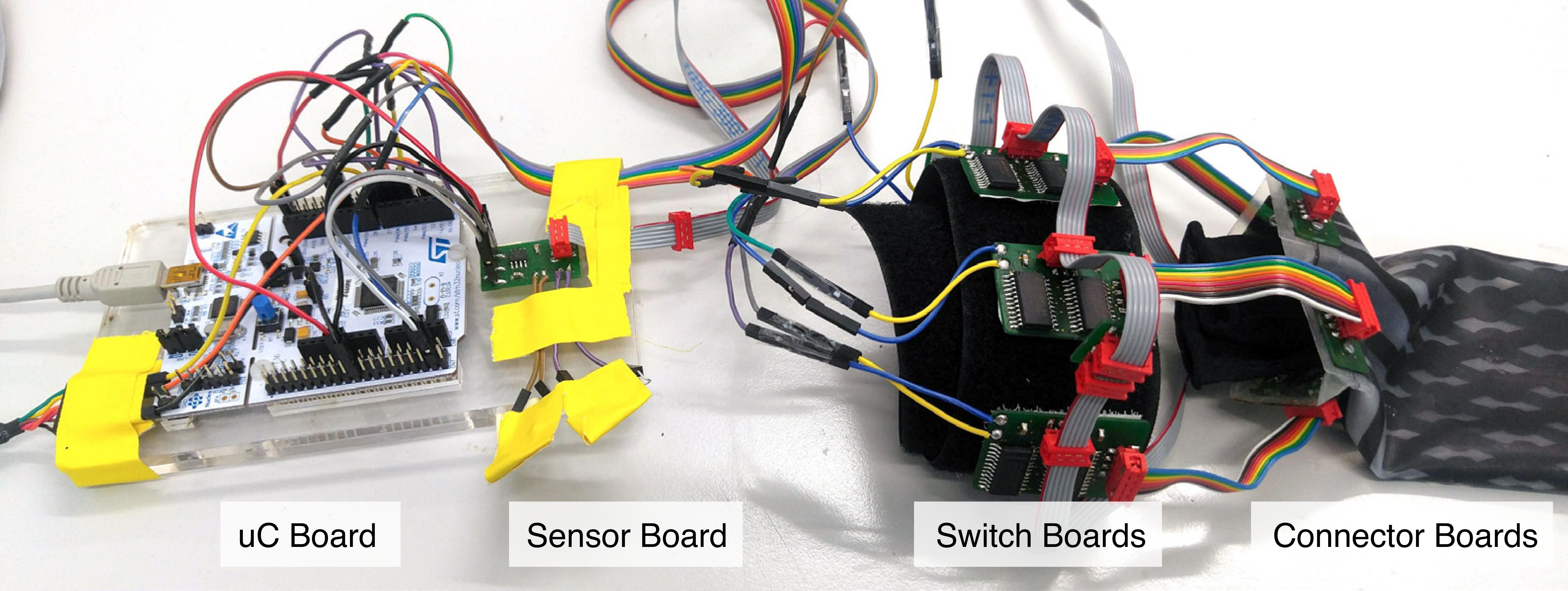}
	\caption{Our custom modular measurement setup with the four types of boards. Up to 8 \emph{switch boards} (and according \emph{connector boards}) can be daisy chained.}
	\label{fig:hardware}
\end{figure}

\label{sec:measuring}
In our setup capacitance is indirectly measured by timing the charging of a capacitor until a predefined voltage level, since the charging time is linearly proportional to the capacitance. However, our setting is more challenging, since we have to dynamically reconnect the electrodes following the measurement protocol described in \secref{sec:array}. For this purpose, we design a modular measuring system (\figref{fig:circuit} right and \figref{fig:hardware}), composed of three kinds of custom boards: the \emph{connector board}, which is directly placed in contact with the sensor, the \emph{switch board}, which is connected to the connector board by a set of flexible wires and the \emph{sensing board} that contains the electronics needed to measure the charging times and send them to the connected computer. The connector boards are placed on the sensor on the exposed sensor pads that are shown in \figref{fig:offsetconnectors}, supported by a PET foil and screwed into an acrylic counter-holder. The PET foil acts as intermediary from stretchable (silicone sensor), through flexible (PET), to fully rigid (\emph{connector board}). The switch boards enable switching through the sensor combinations and they can be daisy-chained to allow for a wide variety of sensor layouts. The switching is controlled from the \emph{uC board}: A STM32 microcontroller on a NUCLEO-F446RE board \cite{STM32Nucleo}. The microcontroller continuously transmits the charging time measurements to the computer via a USB-serial connection.

\begin{figure}
	\centering
	\includegraphics[width=1\linewidth]{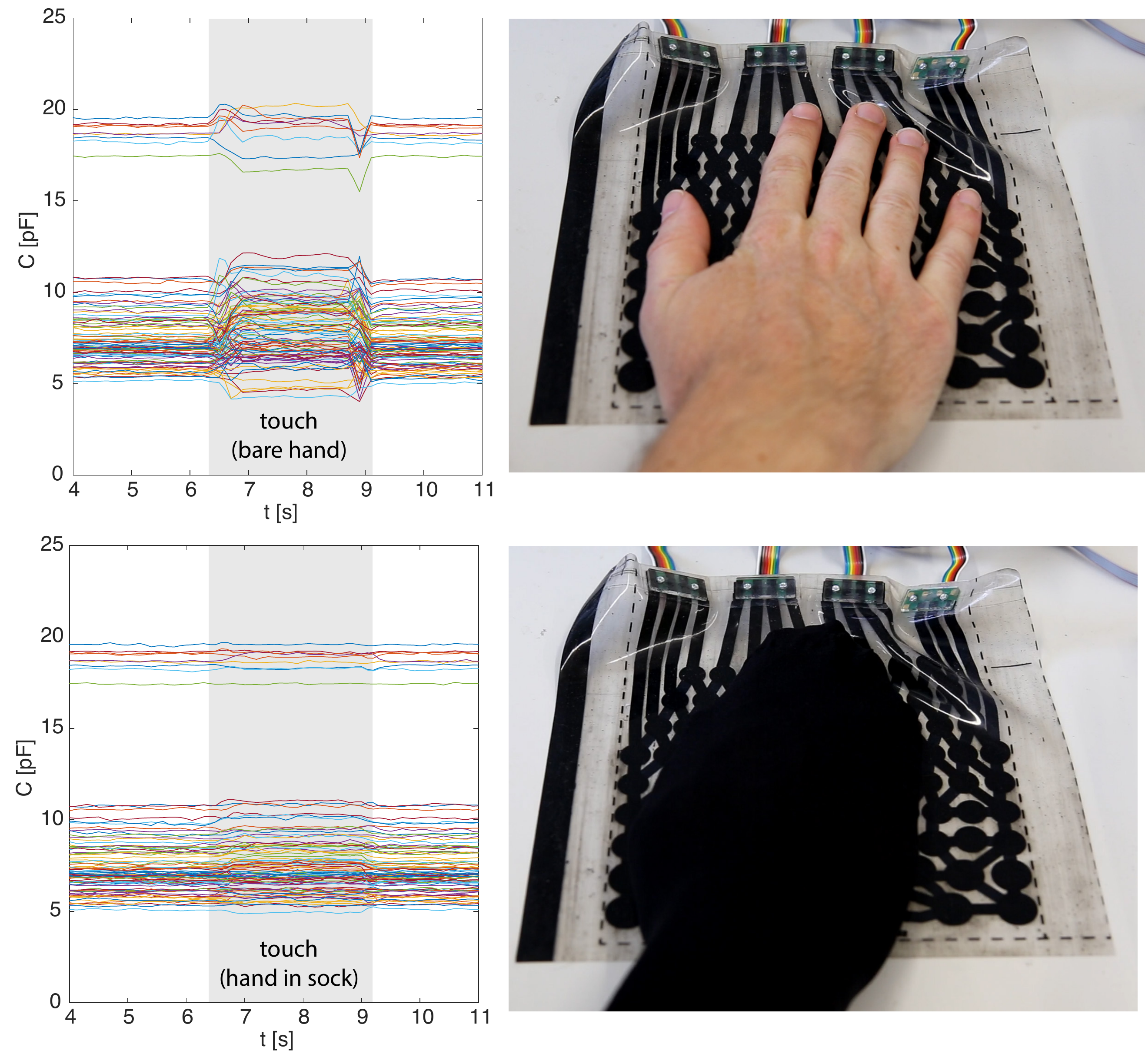}
	\caption{This experiment demonstrates the effect of the nylon sock, worn below the sensor. Top: If the sensor is touched without the sock, the influence of the body capacitance creates clear spikes in the capacitance measured per sensor cell. Bottom: If the nylon sock is worn the same effect is minimal.}
	\label{fig:touching}
\end{figure}

The capacitance measuring circuit (\figref{fig:circuit} left) is implemented using a NE555 timer IC. It outputs a square wave SIGNAL with a frequency $f$ which is converted to capacitance by $C=1/(f\cdot(R_1 + 2 R_2)\cdot \ln(2))$, where $R_1$ and $R_2$ are the charging resistors. The larger these charging resistors are, the slower the capacitors are charged and dis-charged and the longer it takes for a complete measuring round (going through all sets of combined electrodes as shown in \figref{fig:example3x3}) and get the local capacitance changes updated. Note that our model neglects the influence of the resistance of the electrodes themselves. The full resistance for the longest electrode strip is about 50 kOhm. We experimentally found that setting  $R1$ = \unit[470]{kOhm} and $R2$  = \unit[47]{kOhm} is a good compromise that produces sufficient accuracy while still supporting an interactive frame rate of \unit[8]{Hz}.  The parasitic capacitance of the circuit has to be subtracted from all the capacitance measurements. This can be simply done by continuously measuring the capacitance between two unconnected connector board pads. A nylon sock is worn below the sensor when capturing human body part deformation. As demonstrated in \figref{fig:touching}, it shields the in silicone embedded capacitor array from body capacitance and lowers the friction between the sensor and the skin, to e.g. pull a cylindrical sensor over a wrist with much less effort.

\end{document}